\documentclass[preprint, sort&compress]{elsarticle}
\usepackage{graphicx}
\usepackage{multirow}
\usepackage{amsmath}
\usepackage{amssymb}
\usepackage{xspace}
\usepackage{gensymb}
\usepackage{hhline}
\usepackage{multirow}
\usepackage{booktabs}
\usepackage{hyperref}
\usepackage{cleveref}
\usepackage{aas_macros}
\usepackage{soul}
\usepackage{xcolor}
\usepackage{array}
\usepackage{adjustbox}
\usepackage{cancel}
\usepackage{fancyhdr}

\usepackage{etoolbox}
\makeatletter
    \patchcmd{\tnotemark}{\ding{73}}{}{}{\@latex@error{Failed to path \string\tnotemark\space for \string\ding{73}}}
    \patchcmd{\tnotetext}{\ding{73}}{}{}{\@latex@error{Failed to path \string\tnotetext\space for \string\ding{73}}}

\newlength{\wdo}
\newcommand{\stroke}[1]{{$#1$}%
\settowidth{\wdo}{${#1}$} {\kern-\wdo}%
\partialvartstrokedint}

\makeatletter
\newcommand{\fancysep}{%
  \@afterindentfalse
  {\begin{center}
    \resizebox{0.8\linewidth}{0.4ex}{{%
        \fontsize{20}{24}\usefont{U}{webo}{xl}{n}{4}}}%
  \end{center}}\@afterheading}
\makeatother

{\vskip 1.0em
\framebox[\columnwidth][r]{%
\begin{minipage}[c]{\columnwidth}%
\vspace{-1.0em}%
#1%
\end{minipage}}}
{\vskip 1.0em}

\def\XXint#1#2#3{{\setbox0=\hbox{$#1{#2#3}{\int}$}
     \vcenter{\hbox{$#2#3$}}\kern-.5\wd0}}

\newcommand{\beq}{\begin{equation}}
\newcommand{\eeq}{\end{equation}}
\newcommand{\beqa}{\begin{eqnarray}}
\newcommand{\eeqa}{\end{eqnarray}}

\newcommand{\mpl}{\ensuremath{m_\textnormal{pl}}\xspace}

\newcommand{\neff}{\ensuremath{N_\textnormal{eff}}\xspace}

\newcommand{\bardens}{\ensuremath{\omega_b}\xspace}
\newcommand{\burst}{{\sc burst}\xspace}

\newcommand{\gstarsi}{\ensuremath{g_{\star S}(\ainit)}\xspace}

\newcommand{\gstarsf}{\ensuremath{g_{\star S}(\afin)}\xspace}

\newcommand{\gstars}{\ensuremath{g_{\star S}}\xspace}

\newcommand{\tin}{\ensuremath{T_\textnormal{cm}(a_\textnormal{in})}\xspace}
\newcommand{\ain}{\ensuremath{a_\textnormal{in}}\xspace}
\newcommand{\ainit}{\ensuremath{a_{i}}\xspace}
\newcommand{\afin}{\ensuremath{a_{f}}\xspace}
\newcommand{\adec}{\ensuremath{a_\textnormal{dec}}\xspace}
\newcommand{\tdec}{\ensuremath{T(a_\textnormal{dec})}\xspace}
\newcommand{\xdec}{\ensuremath{x(a_\textnormal{dec})}\xspace}

\newcommand{\tcm}{\ensuremath{T_\textnormal{cm}}\xspace}

\newcommand{\tcmpl}{\ensuremath{T_\textnormal{cm}/T}\xspace}
\newcommand{\tcmplfo}{\ensuremath{[T_\textnormal{cm}/T]_\textnormal{f.o.}}\xspace}
\newcommand{\np}{\ensuremath{n/p}\xspace}

\newcommand{\nue}{{\ensuremath{\nu_{e}}}\xspace}

\newcommand{\bnue}{\ensuremath{\overline{\nu}_e}\xspace}

\newcommand{\spl}{\ensuremath{s_\textnormal{pl}}\xspace}
\newcommand{\snu}{\ensuremath{s_\nu}\xspace}
\newcommand{\stot}{\ensuremath{s_\textnormal{tot}}\xspace}

\newcommand{\ben}{\begin{enumerate}}
\newcommand{\een}{\end{enumerate}}

\newcommand{\heiv}{\ensuremath{\,^4\textnormal{He}}\xspace}

\newcommand{\yp}{\ensuremath{Y_P}\xspace}

\newcommand{\felev}{\ensuremath{[4/11]^{1/3}}\xspace}
\newcommand{\tstart}{\ensuremath{T(a_s)}\xspace}
\newcommand{\xstart}{\ensuremath{x(a_s)}\xspace}

\newcommand{\fhat}{\ensuremath{\hat{F}}\xspace}

\begin{document}

\title{Insights into neutrino decoupling gleaned from considerations of the
role of electron mass}

\author{E.~Grohs\corref{cor1}}
\ead{egrohs@umich.edu}
\address{Department of Physics, University of Michigan, Ann Arbor, Michigan
48109, USA}
\cortext[cor1]{Corresponding author}

\author{George M.~Fuller\corref{cor2}}
\ead{gfuller@ucsd.edu}
\address{Department of Physics, University of California,
San Diego, La Jolla, California 92093, USA}

\date{\today}

\begin{abstract} 

We present calculations showing how electron rest mass influences entropy flow,
neutrino decoupling, and Big Bang Nucleosynthesis (BBN) in the early universe.
To elucidate this physics and especially the sensitivity of BBN and related
epochs to electron mass, we consider a parameter space of rest mass values
larger and smaller than the accepted vacuum value.  Electromagnetic
equilibrium, coupled with the high entropy of the early universe, guarantees
that significant numbers of electron-positron pairs are present, and dominate
over the number of ionization electrons to temperatures much lower than the
vacuum electron rest mass.  Scattering between the electrons-positrons and the
neutrinos largely controls the flow of entropy from the plasma into the
neutrino seas.  Moreover, the number density of electron-positron-pair targets
can be exponentially sensitive to the effective in-medium electron mass.  This
entropy flow influences the phasing of scale factor and temperature, the
charged current weak-interaction-determined neutron-to-proton ratio, and the
spectral distortions in the relic neutrino energy spectra.  Our calculations
show the sensitivity of the physics of this epoch to three separate effects:
finite electron mass, finite-temperature quantum electrodynamic (QED) effects
on the plasma equation of state, and Boltzmann neutrino energy transport.  The
ratio of neutrino to plasma-component energy scales manifests in Cosmic
Microwave Background (CMB) observables, namely the baryon density and the
radiation energy density, along with the primordial helium and deuterium
abundances.  Our results demonstrate how the treatment of in-medium electron
mass (i.e., QED effects) could translate into an important source of
uncertainty in extracting neutrino and beyond-standard-model physics limits
from future high-precision CMB data.

\end{abstract}


\begin{keyword}
Big Bang Nucleosynthesis\sep Weak Interactions \sep Cosmological Neutrinos
\sep Early Universe \sep arXiv: 1706.03391
\end{keyword}


\maketitle

\section{Introduction}\label{sec:intro}

In this paper we dissect a key aspect of the physics operating in the epoch of
the early universe where neutrinos cease to efficiently exchange energy with
the photon-electron-positron plasma and evolve into freely falling particles.
By unphysically varying the bare vacuum mass of the electron we can gain
insight into the timelike entropy flow from the plasma into the decoupling
neutrino component and the evolution of the neutron-to-proton ratio through the
epoch of Big Bang Nucleosynthesis (BBN).  We do not consider the timelike
variation of the electron rest mass as possible beyond-standard-model physics
in this work.  By keeping all other fundamental constants fixed, we merely
consider the role the electron mass has during BBN.  See Ref.\
\cite{2014arXiv1412.1078S} for time variation of fundamental constants and
particle masses.

The early universe possesses high entropy-per-baryon (alternatively low baryon
to photon ratio $\eta=6.1\times10^{-10}$) as derived from the Cosmic Microwave
Background (CMB) anisotropies \cite{PlanckXIII:2015} and as independently
inferred from the primordial deuterium abundance
\cite{2003ApJS..149....1K,2016ApJ...830..148C} and BBN calculations.  The high
entropy and concomitant disorder set the stage for the key features of neutrino
decoupling and BBN.

In particular, high entropy in the early universe implies significant
equilibrium electron-positron pair densities, even at temperatures well below
the electron rest mass.  Neutrino-electron/positron scattering in large part
controls the exchange of energy and entropy between the plasma and neutrino
components.  In turn, the equilibrium number densities of electron-positron
pairs can be exponentially sensitive to the electron mass, especially in the
later stages of neutrino decoupling and BBN.

Though the baryon rest mass density is low, e.g., $\sim10^{-4}\,{\rm g/cm}^{3}$
at a temperature $\sim1\,{\rm keV}$, the early universe is nevertheless a
polarizable, high energy density, relativistic plasma.  The effective in-medium
masses of the electron and positron are the relevant determinants of the number
density of neutrino scattering targets.  In analogy with the vacuum case, we
can take the form of the in-medium dispersion relation to be $E = \sqrt{p^2 +
m_{\rm eff}^2}$ for three-momentum magnitude $p$.  $m_{\rm eff}$ represents the
effective mass of an electron or positron in medium.  We will discuss how this
term can differ from the bare vacuum mass $m_e$ as a function of temperature,
momentum, and vacuum electron mass itself.  There are many studies of the
dispersion relations for the electron, positron, and also photon
quasi-particles in this medium.  These quantum electrodynamic (QED) effects or
``plasma corrections''
\cite{1982NuPhB.209..372C,1994PhRvD..49..611H,1997PhRvD..56.5123F,1999PhRvD..59j3502L}
are commonly employed in the more sophisticated treatment of BBN (see for
example, Refs.\ \cite{neff:3.046,2008CoPhC.178..956P,transport_paper}).

Observational cosmology is entering an era of two distinct, high-precision
measurements: CMB observables, e.g., the effective number of relativistic
degrees of freedom, \neff, and the primordial helium abundance, \yp; and
high-redshift astronomical observables, e.g., the primordial deuterium
abundance.  The proper interpretation of these measurements necessitates
detailed calculations of the epochs surrounding primordial nucleosynthesis.  By
examining the role of electron mass in these calculations, we are led to
conclude that the in-medium electron/positron mass corrections take on a new
and heightened significance.

We use natural units $\hbar=c=k_B=1$ and assume neutrinos are massless
throughout this paper.  In Section \ref{sec:therm}, we discuss the
statistical mechanics and thermodynamics particular to the early universe.  We give
an exposition of how the vacuum electron rest mass perturbs the early universe
thermodynamics in Sec.\ \ref{sec:notrans}.  Section \ref{sec:qed} discusses
the finite-temperature QED corrections to the plasma equation of state.  In
Sec.\ \ref{sec:trans}, we describe the effect of electron rest mass on
Boltzmann neutrino transport and nuclear reactions.  We give conclusions in
Sec.\ \ref{sec:concl}.  A note on our notation: we will predominantly use the
scale factor $a$ as the independent variable in our equations.  For the sake of
clarity, we will denote dependent variable values at specific epochs as $Q(a)$
where the quantity $Q$ is a function of $a$.

\section{Overview of statistical mechanics in the early universe}\label{sec:therm}

A salient feature of the evolution of the early universe is that the Hubble
expansion rate, driven by gravitation, is inherently slow.  An inevitable
result of the slow expansion is to conduce strong and electromagnetic
interactions (and even the weak interaction at substantially high temperatures)
to maintain the constituents of the early universe in thermal and chemical
equilibrium.  Eventually the weak interaction is
not strong enough to maintain equilibrium between the photon-electron-positron
plasma and the neutrino seas.  This event, sometimes termed weak decoupling,
occurs nearly simultaneously with the epoch where the nuclear reactions ---
strong, electromagnetic and weak --- also drop out of equilibrium.
Nevertheless, the electron, positron, and photon components of the plasma (with
the thermal coupling to baryons) remain well described by equilibrium
thermodynamics over the vast range of BBN epochs.  In thermal and chemical
equilibrium, along with a homogeneous and isotropic geometry, the comoving
entropy is conserved \cite{1990eaun.book.....K}.

\subsection{Comoving temperature parameter}

We will focus on high-precision calculations of ratios involving photons,
neutrinos, and baryons.  At early times/high temperatures $(T\gtrsim10\,{\rm
MeV})$, the neutrinos are thermally and chemically coupled to the
electromagnetic plasma of photons, electrons, positrons, and a small abundance
of baryons.  Eventually, the neutrinos decouple from both the plasma and
baryons and free stream.  The epoch of neutrino decoupling roughly ceases at
$T\sim10\,{\rm keV}$.  In order to calculate the energy density in radiation,
we need to consider two scales.  First, the plasma temperature $T$ dictates the
energy density in photons.  As the universe expands, the plasma temperature
decreases while the scale factor, $a$, increases.  The product of $T$ with $a$
is not a comoving invariant during the BBN epoch.  The relic positrons
annihilate with electrons to form photons.  These photons scatter on the
remaining charged leptons and quickly thermalize. If the photons have a high
enough energy, they can create an electron-positron (denoted $e^\pm$) pair,
keeping the charged leptons in thermal and chemical equilibrium with the
photons.  As the universe expands and cools, the thermally distributed photons
do not have the energies required to pair create $e^\pm$.  The result is that
the heat in the form of charged leptons is transferred to photons.  As this all
occurs in thermal equilibrium, there is no change in the entropy within the
plasma and so the product $Ta$ must increase.  The elimination of statistical
degrees of freedom is not the only way that the product $Ta$ can increase.  The
entropy density is proportional to the product of the number of degrees of
freedom and the cube of the temperature.  If the entropy rises, and the number
of degrees of freedom remains the same, then the temperature will rise.  We
will consider both mechanisms in this paper.

To compare with the plasma temperature, we will introduce another energy scale
called the comoving temperature parameter, \tcm, whose product with $a$ is a
comoving invariant
\beq\label{eq:tcm}
  \tcm(a)=\tin\left[\frac{a_{\rm in}}{a}\right],
\eeq
where we have written \tcm as a function of $a$.  \tin is the plasma
temperature at an initial epoch of our choosing, which we label \ain.  We
normally would choose \ain at an early enough epoch so that the neutrinos are
in thermal equilibrium with the charged leptons.  As neutrino decoupling
proceeds, the neutrinos maintain occupation numbers close to Fermi-Dirac (FD)
equilibrium with a temperature-like parameterization close to \tcm.  In one
sense, \tcm can be considered a neutrino temperature.  We caution against this
interpretation as the neutrinos are no longer in thermal equilibrium with each
other at late times and so the strict thermodynamic definition of temperature
is not an applicable quantity.  Nevertheless, the neutrinos have an energy
density and that energy density is described by the \tcm scale.

\subsection{Standard equilibrium value of the temperature ratio}\label{ssec:theory}

We are interested in epochs of the early universe where the energy density (and
by extension entropic density) is dominated by radiation. To begin, we give the
standard explanation of how to calculate the ratio of $\tcm/T$ at freeze-out --
these arguments can be found in many textbooks (see Refs.\
\cite{1990eaun.book.....K,Dodelson:2003mc,2008cosm.book.....W}).  To compare
with the more sophisticated treatments in Secs.\ \ref{sec:notrans} --
\ref{sec:trans}, we first give the equilibrium argument.  We will carefully
consider the assumptions the equilibrium argument rely on later.

\tcm and $T$ give the two energy scales for neutrinos and photons,
respectively.  We can characterize the difference in scales using entropy
arguments.  We describe the entropy, $S$, of an ideal gas using its extensive
property \cite{Pathria_stat_mech}:
\beq\label{intro-eqn:entropydef}
  S=\frac{E+PV-\sum\mu_iN_i}{T},
\eeq
where $E$ is the total internal energy, $P$ is the pressure, $V$ is the volume,
$\mu_i$ is the chemical potential for a given species $i$, and $N_i$ is the
number of particles of species $i$.  We will assume the chemical potentials are
small for the plasma constituents which carry the bulk of the entropy at any
given time.  If we write the total internal energy in terms of the energy
density and the volume, $E=\rho V$, we can rearrange terms to find the entropic
density
\beq
  S/V = \frac{\rho+P}{T},\label{eq:ent_dens_def}
\eeq
in terms of the energy density, pressure, and temperature.  Very simply, since
the entropy is conserved in equilibrium, we can relate the initial $(i)$
temperature and the final $(f)$ temperature as follows
\beq\label{eq:tratio1}
  S(\ainit)=S(\afin)\implies\frac{T(\afin)}{T(\ainit)}=
  \left\{\frac{[S/V](\ainit)\,\frac{\ainit^3}{T^3(\ainit)}}
  {[S/V](\afin)\,\frac{\afin^3}{T^3(\afin)}}\right\}^{1/3},
\eeq
where we have used the fact that $V\propto a^3$.  For radiation, the energy
density scales as $\rho\propto gT^4$, and the equation of state is $P=\rho/3$.
The statistical weight in relativistic particles is $g$.
From Eq.\
\eqref{eq:ent_dens_def}, we are left with an expression for the entropic
density $S/V\propto gT^3$.  If our gas has multiple components with differing
temperatures and statistical weights $g$, we can write the entropic density as
\begin{align}
  S/V &= \frac{2\pi^2}{45}T^3\left\{\sum\limits_mg_m\left[\frac{T_m}{T}\right]^4
  +\frac{7}{8}\sum\limits_ng_n\left[\frac{T_n}{T}\right]^4\right\}\\
  &\equiv\frac{2\pi^2}{45}\gstars T^3,\label{eq:gstars}
\end{align}
where \gstars is an effective-entropic-spin statistic
\cite{1990eaun.book.....K}.  The sum over $m$ is for bosonic species, and the
sum over $n$ is for fermionic species.  In the definition of \gstars, we allow
for the plasma constituents to have different temperatures or temperature
parameters, although this expressions rests on assumed Fermi-Dirac or
Bose-Einstein shaped energy spectra.

Using the expression in Eq.\ \eqref{eq:gstars}, we can solve for the ratio of
temperatures in Eq.\ \eqref{eq:tratio1}
\beq\label{eq:tratio2}
  \frac{T(\ainit)}{T(\afin)}\frac{a_i}{a_f} =
  \left[\frac{\gstarsf}{\gstarsi}\right]^{1/3}.
\eeq
If we pick an initial epoch in Eq.\ \eqref{eq:tratio2} to match our choice for
the initial epoch in Eq.\ \eqref{eq:tcm}, i.e., $T(\ainit)=\tin$, we can write
$T(\ainit)=T_{{\rm cm}}(\afin)[a_f/a_i]$ to find
\beq\label{eq:tratio3}
  \frac{\tcm(\afin)}{T(\afin)} = \left[\frac{\gstarsf}{\gstarsi}\right]^{1/3}.
\eeq
Given the parameterization of \gstars in Eq.\ \eqref{eq:gstars}, Eq.\
\eqref{eq:tratio3} is true at any temperatures so long as the particles have
equilibrium-shaped energy distribution functions.

Completely decoupled neutrinos have a fixed product of scale factor and
neutrino temperature parameter, reflecting how the three-momentum magnitude of
a free-falling neutrino redshifts with inverse scale factor.  If neutrinos were
taken to decouple instantaneously, then their energy and momentum distribution
functions at that epoch would have a FD-shaped energy distribution
characterized by two quantities: the chemical potential, and the temperature
parameter.  That temperature parameter is the same as the definition of \tcm in
Eq.\ \eqref{eq:tcm}.  Using Eq.\ \eqref{eq:tratio3} and taking $T(a_f)$ after
the electron-positron annihilation epoch, we have
\beq\label{eq:tratio4}
  \left[\frac{\tcm}{T}\right]_{\rm f.o.} = \left[\frac{4}{11}\right]^{1/3},
\eeq
where we have dropped the $a_f$ arguments and replaced them with a
``freeze-out'' subscript f.o.  In deriving this value, we have made the
following assumptions:
\ben
  \item the neutrinos, independent of energy, decouple with a FD shaped
  energy-distribution function and a temperature parameter synchronized with the
  plasma temperature at an epoch designated by \adec;\label{list:m_e}
  \item the contribution at $T(\adec)$ to \gstars from the charged leptons is
  $(7/8)\times4$, which neglects effects of  a nonzero value of $m_e$;
  \item the comoving entropy in the plasma is conserved;\label{list:ds0}
  \item finite-temperature QED effects on the equation of state
  for electrons, positrons, and photons are negligible;\label{list:ft}
  \item the electrons and positrons have negligible chemical
  potentials.\label{list:mu}
\een
In this paper, we will evaluate the sensitivity of \tcmplfo to items
\ref{list:m_e} -- \ref{list:ft}.  We will not consider how \tcmplfo changes
with item \ref{list:mu}, although we will investigate how the chemical
potential affects the $e^\pm$ pair density.

\subsection{Pair density of electrons and positrons}

The number density for electrons or positrons is
\beq
  n_{e^\pm} = \frac{g}{[2\pi]^3}\int_0^{\infty}d^3p
  \frac{1}{e^{[E\pm\mu_e]/T}+1},
\eeq
where $d^3p$ is the momentum phase-space density, $E$ is the energy, and
$\mu_e$ is the chemical potential of the {\it electron}.  We assume chemical
equilibrium between the electron and positron seas, i.e.,
$\mu_e\equiv\mu_{e^-}=-\mu_{e^+}$, and take $g=2$, implying
\begin{align}
  n_{e^-}-n_{e^+} &= \frac{1}{\pi^2}\left\{
  \int_0^{\infty}dp\frac{p^2}{e^{[E-\mu_e]/T} + 1}
  -\int_0^{\infty}dp\frac{p^2}{e^{[E+\mu_e]/T} + 1}
  \right\}\label{eq:nep_diff}\\
  &\approx \frac{T^3}{6\pi^2}\left\{\pi^2\left[
  \frac{\mu_e}{T}\right] + \left[\frac{\mu_e}{T}\right]^3\right\},
\end{align}
where the last approximation assumes temperatures high enough that electrons
have extreme relativistic kinematics.  If we restrict ourselves to epochs where
the only charge-carrier constituents of the universe are electrons, positrons,
and protons, then the left hand side of Eq.\ \eqref{eq:nep_diff} is equal to
the number density of protons by charge neutrality
\beq
  n_{e^-}-n_{e^+} = n_p.
\eeq
We term the excess of electrons over positrons ``ionization electrons'', which
are equal to the number of protons.  The number density of protons is much
smaller than the number density of a plasma particle (photons and $e^\pm$).  As
very high temperature $(T>>m_e)$, we can ignore the cubic term on the
right-hand side of Eq.\ \eqref{eq:nep_diff} to show $\mu_e/T\propto
n_p/T^3\sim\eta$, where $\eta\simeq 6\times10^{-10}$ is the baryon-to-photon
ratio.

\begin{figure}
  \includegraphics[width=\columnwidth]{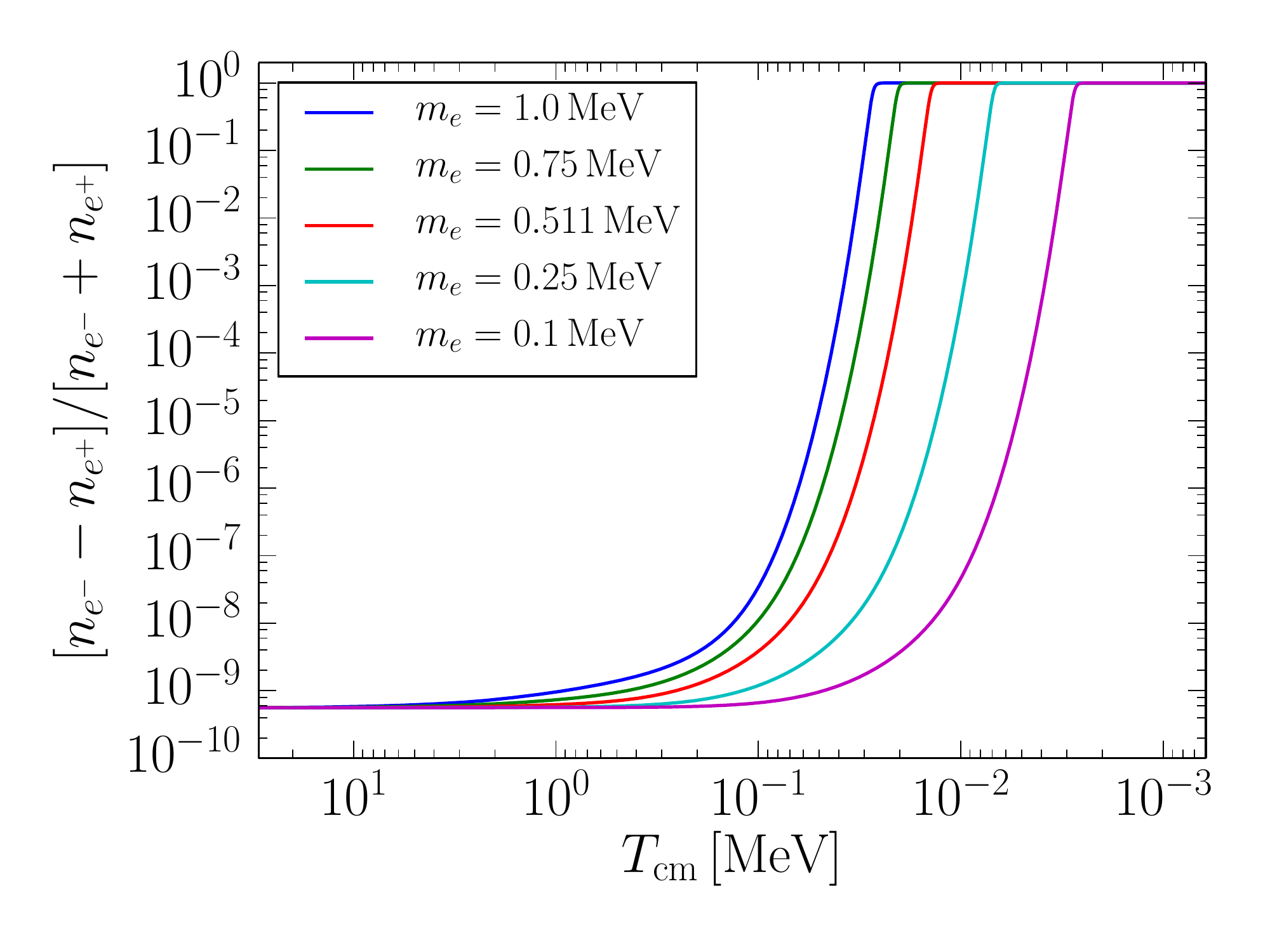}
  \caption{\label{fig:nep_vs_tcm} The difference in electron and positron
  number densities normalized by the sum of $e^\pm$ number densities, plotted
  against \tcm.  The five curves correspond to five different assumed values of
  the vacuum electron rest mass, $m_e$.}
\end{figure}

Figure \ref{fig:nep_vs_tcm} shows the ratio of number of ionization electrons
(equal to the difference $n_{e^-}-n_{e^+}$) to total charged leptons (equal to
the sum $n_{e^-}+n_{e^+}$) versus the comoving temperature parameter for five
different assumed electron vacuum rest-mass values.  The third value, denoted
$m_e=0.511\,{\rm MeV}$, is the true value of the electron vacuum rest mass.  At
all temperatures, the number of ionization electrons is equal to the number of
protons which is the same order of magnitude as the baryon number.  At high
temperature, the total number of charged leptons is the same order of magnitude
as the number of photons.  Therefore, for $\tcm>>1\,{\rm MeV}$, all five curves
should converge to a value similar to $\eta$.  Furthermore, as we assume all
positrons eventually annihilate with electrons, all five curves will converge
to unity at low temperatures.  Changing the mass changes when the pairs
disappear, or equivalently, when the epoch of $e^\pm$ annihilation
occurs.  All five curves in Fig.\ \ref{fig:nep_vs_tcm} show that the number of
pairs (equal to $n_{e^+}$) dominate over the number of ionization electrons
until late times, specifically $\tcm<<m_e$.

\section{Changing $m_e$ in the instantaneous weak decoupling scenario}
\label{sec:notrans}

In order to calculate how a nonzero $m_e$ changes the ratio \tcmpl, we must
start with Eq.\ \eqref{eq:tratio1} and calculate the changes to the entropy.
Equation \eqref{eq:tratio3} no longer applies since the charged lepton energy
density is not proportional to the fourth power of temperature, i.e.,
$\rho\propto gT^4$.

In the case of a non-degenerate ($\mu=0$) fermionic species, the energy density
for an ideal gas at temperature $T$ is
\beq
  \rho = \frac{g}{[2\pi]^3}\int d^3p\frac{E(p)}{e^{E(p)/T} + 1}.
  \label{eq:rho_def}
\eeq
If the particles are massless, then the dispersion relation is $E(p)=p$, and the
energy density reduces to \beq
  \rho^{(m=0)}=\frac{7\pi^2g}{240}T^4.
\eeq
If the particles have small nonzero masses, i.e.\ $m<<T$, the dispersion
relation is \begin{align}
  E(p)&=\sqrt{p^2+m^2}\\
  &\simeq p + \frac{m^2}{2p},\label{eq:disprelapprox}
\end{align}
to second order in $m$.  Substituting Eq.\ \eqref{eq:disprelapprox} into
Eq.\ \eqref{eq:rho_def}, we find
\beq\label{eq:rho_m}
  \rho^{(m<<T)} = \rho^{(m=0)}\left\{1-\frac{5}{7\pi^2}
  \left[\frac{m}{T}\right]^2\right\},
\eeq
where we have only kept terms to order $m^2$.  We will define $x$ such that
$x\equiv m/T$.  We find a similar expression to Eq.\ \eqref{eq:rho_m} for the
pressure
\beq\label{eq:p_m}
  P = P^{(m=0)}\left[1 - \frac{15}{7\pi^2}x^2\right],
\eeq
and the entropic density
\beq\label{eq:sv_m}
  S/V = (S/V)^{(m=0)}\left[1-\frac{15}{14\pi^2}x^2\right],
\eeq
where we have dropped the superscript label $(m<<T)$ on the left-hand-side of
Eqs.\ \eqref{eq:p_m} and \eqref{eq:sv_m} for ease in notation.  If we
re-examine the statistics of $e^\pm$ annihilation, but drop the
assumption that $m_e=0$, then conservation of comoving entropy implies
\beq
  \frac{2\pi^2}{45}T^3(\ainit)\ainit^3
  \left\{2 + \frac{7}{8}[2+2]\left[
  1-\frac{15}{14\pi^2}x^2(\ainit)\right]\right\}
   = \frac{2\pi^2}{45} T^3(\afin)\afin^3\left\{
  \vphantom{\frac{15}{14\pi^2}}2\right\},\label{eq:scons_me}
\eeq
where we have written $x$ at the initial epoch as a function of \ainit.  The
curly brackets in the left-hand-side of Eq.\ \eqref{eq:scons_me} show the
change in \gstarsi to second order in $m_e$, whereas the right-hand-side is
simply the case when only photons contribute to the entropy.  Solving for the
ratio of temperatures in Eq.\ \eqref{eq:scons_me} yields
\beq\label{eq:tcmpl_z}
  \left[\frac{\tcm}{T}\right]_{\rm f.o.} 
  = \left[\frac{4}{11}\right]^{1/3}
  \left[1+\frac{5}{22\pi^2}x^2(\ainit)\right].
\eeq

\begin{figure}
  \includegraphics[width=\columnwidth]{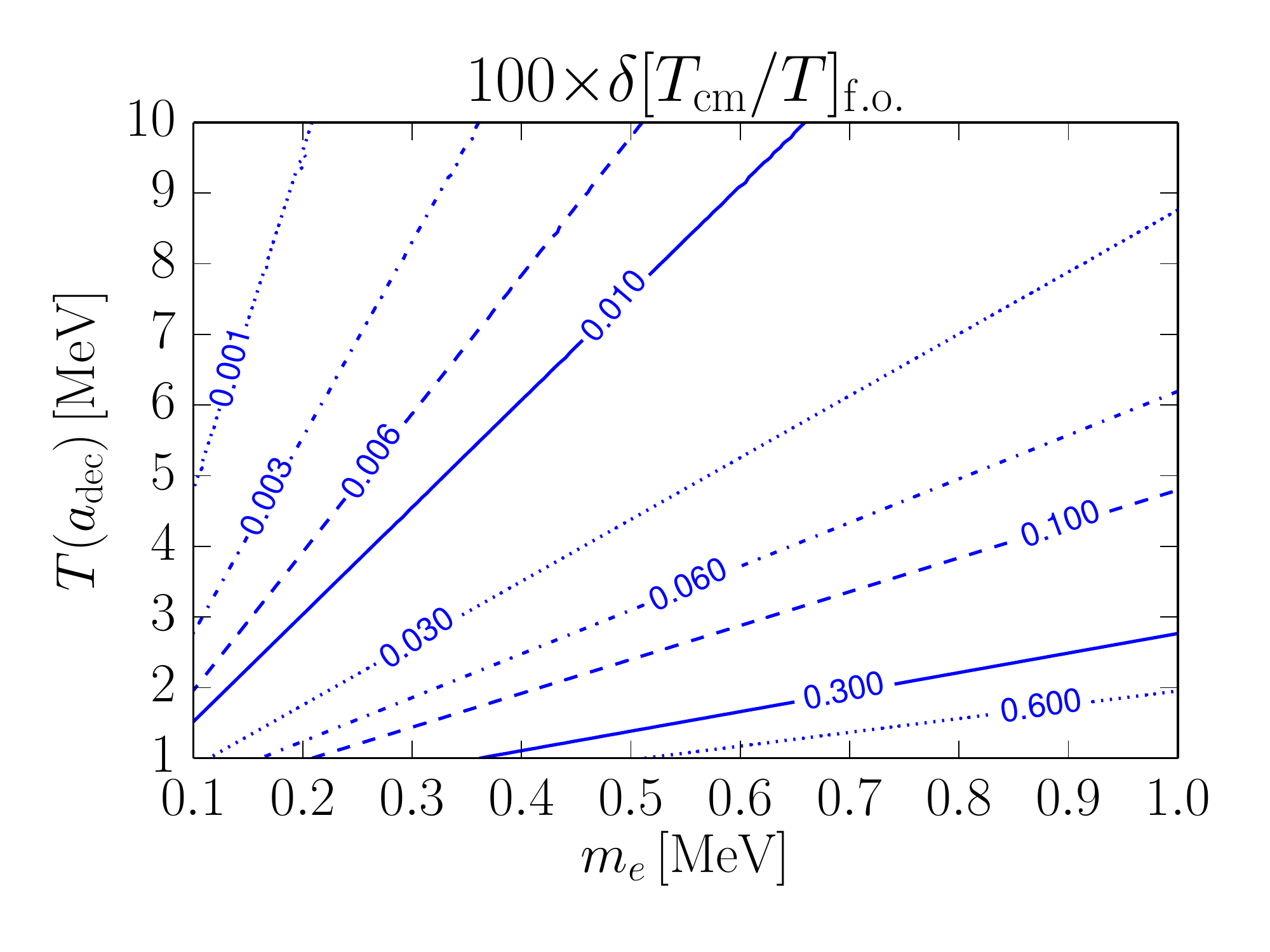}
  \caption{\label{fig:tdec_vs_xm_contour_tcmpl} Contours of
  $100\times\delta\tcmplfo=100\times\{\tcmplfo - \felev\}/\felev$ are plotted in
  the assumed electron rest mass $m_e$ -- \tdec plane.  The contours are
  calculated using the full \burst code, however, they agree with the analytic
  estimate in Eq.\ \eqref{eq:tcmpl_z}.}
\end{figure}

We will write $x(\ainit)$ as $\xdec=m_e/\tdec$ in accordance with item
\ref{list:m_e} of the list in Sec.\ \ref{ssec:theory}.  Figure
\ref{fig:tdec_vs_xm_contour_tcmpl} shows contours of constant
$100\times\delta\tcmplfo$ in the \tdec versus $m_e$ parameter space, where we
take
\beq\label{eq:reldiff}
  \delta\tcmplfo\equiv\frac{\tcmplfo - [4/11]^{1/3}}{[4/11]^{1/3}}.
\eeq
The curves in Fig.\ \ref{fig:tdec_vs_xm_contour_tcmpl} were calculated using
our code \burst \cite{transport_paper}.  The contour locations agree to high
precision with Eq.\ \eqref{eq:tcmpl_z} over the entire parameter space.  The
agreement is the best for small \xdec and slightly degrades for increasing
$m_e$ and decreasing \tdec as expected from Eq.\ \eqref{eq:tcmpl_z}.  As \tdec
decreases, there are fewer $e^\pm$ pairs remaining once the neutrinos decouple.
The result is that the photons do not heat up as much as when \tdec is large,
and so \tcmplfo increases.  Similarly, as $m_e$ increases, there is a smaller
energy density of $e^\pm$ pairs and \tcmplfo increases by the same logic as the
\tdec dependence.  For a given contour value, Eq.\ \eqref{eq:tcmpl_z} states
$\tdec\propto m_e/\tcmplfo^{1/2}$, implying that the slope of the contour will
increase with decreasing \tcmplfo.  The contour where $\delta\tcmplfo$ is
identically zero is reached if $m_e$ is set to zero.

Figure \ref{fig:tdec_vs_xm_contour_tcmpl} shows that \tcmplfo is always
slightly larger than \felev.  Although this is useful for calculating the
energy density of neutrinos given the plasma temperature, \tcm and \tcmpl
themselves are not physical observables.  We can use the baryon number density
as another physical observable to tease out the value of \tcmplfo.  To
accomplish this task, we will utilize the {\it baryon density} \bardens.
\bardens is related to the contribution of baryon rest mass to the closure
density of the universe, $\Omega_b$, and the Hubble parameter, $h$
\beq
  \bardens=\Omega_bh^2,
\eeq
where $h$ is used to parameterize the Hubble expansion rate at the current
epoch, $H_0$
\beq
  H_0 = 100\times h\,\,{\rm km/s/Mpc}.
\eeq
By using \bardens, we can calculate the proper baryon number density.  We can
relate the baryon number density at the current epoch $n_b(a_0)$ to the baryon
number density at any epoch by the following
\beq\label{eq:bn1}
  n_b(a_0) = n_b(a)\left[\frac{a}{a_0}\right]^3,
\eeq
where we have assumed the product of number density and the cube of the scale
factor is a comoving invariant.  $a_0$ is the scale factor at the current
epoch.  The baryon density at the current epoch is then measured by CMB
experiments using the baryon density, \bardens
\beq\label{eq:bn2}
  n_b(a_0) = \frac{3\mpl^2}{8\pi m_b}\bardens\times[10^2\,{\rm km/s/Mpc}]^2,
\eeq
where \mpl is the Planck mass and $m_b$ is the baryon rest mass.  In this work,
we will use $\bardens=0.022068$ from Ref.\ \cite{PlanckXVI:2014} which is
identical within statistical precision to the updated value in Ref.\
\cite{PlanckXIII:2015}.  To calculate the baryon number density at any epoch,
we can use Eqs.\ \eqref{eq:bn1}, \eqref{eq:bn2}, and our definition for \tcm
\beq\label{eq:bn3}
  n_b(a) = \frac{3\mpl^2}{8\pi m_b}\bardens
  \left[\frac{\tcm(a)}{T_{\rm cm}(a_0)}\right]^3
  \times[10^2\,{\rm km/s/Mpc}]^2.
\eeq
As \tcm is a construct and not a physical observable, we must write the ratio
of comoving temperature parameters in Eq.\ \eqref{eq:bn3} in terms of plasma
temperatures.  As a zeroth approximation for our purposes, we use Eq.\
\eqref{eq:tratio4} to write
\beq\label{eq:bn4}
  n_b^{(z)}(a) = \frac{3\mpl^2}{8\pi m_b}\bardens\frac{11}{4}
  \left[\frac{T(a)}{T(a_0)}\right]^3\times[10^2\,{\rm km/s/Mpc}]^2.
\eeq
We adorn $n_b$ with a superscript $(z)$ to denote that Eq.\ \eqref{eq:bn4} is a
zeroth approximation since we have ignored the contribution from the nonzero
electron rest mass.  When we run \burst with a finite nonzero value of $m_e$,
we find slight discordance between our calculated value of $n_b(a_0)$ and that
of the true value in Eq.\ \eqref{eq:bn2}.  To correct for the discrepancy, we
run another iteration of \burst with a corrected baryon number density
\beq\label{eq:bn5}
  n_b(a) = \frac{3\mpl^2}{8\pi m_b}\bardens\frac{11}{4}\mathcal{C}
  \left[\frac{T(a)}{T(a_0)}\right]^3\times[10^2\,{\rm km/s/Mpc}]^2,
\eeq
where $\mathcal{C}$ is our correction factor.  It is possible to analytically
calculate an estimate of $\mathcal{C}$ using entropy conservation.  However,
there is a slight subtlety we need to address to do so.

We can incorporate the baryon density into our previous nomenclature if we
depart from using the entropic density and instead use the ratio of entropic
density to baryon number density
\beq\label{eq:s_def}
  s\equiv\frac{S/V}{n_b}
\eeq
which we will call the entropy per baryon.  The comoving invariant quantity,
$[S/V]a^3$, we first employed in Eq.\ \eqref{eq:tratio1} becomes the entropy
per baryon quantity $s$.  Up until this point, we have used entropy
conservation \textit{in the plasma} (or equivalently conservation of entropy
per baryon in the plasma) to calculate ratios of quantities before and after
certain epochs.  As we vary both $m_e$ and \tdec, we are changing the entropy
per baryon in the plasma at the epoch \adec via Eq.\
\eqref{eq:sv_m}.  Because the entropy per baryon in the plasma is proportional to
the quotient of plasma temperature cubed to baryon number density, a different
plasma entropy per baryon at neutrino decoupling is equivalent to a different
baryon density at the current epoch.  Therefore, entropy conservation in the
plasma is not the relevant quantity to investigate.  Alternatively, we will
consider the total entropy of the universe, i.e., the sum of the plasma and
neutrino components
\begin{align}
  \stot& = \spl + \snu\\
  &=\frac{1}{n_b}\left[\frac{\rho+P}{T}\right]_{\rm pl} +
  \frac{[S/V]_\nu}{n_b}.
\end{align}
Neutrinos will thermodynamically decouple from the plasma, implying that we
cannot use Eqs.\ \eqref{eq:ent_dens_def} and \eqref{eq:s_def} to determine the
entropy per baryon in the neutrino seas in general.  However, at this point we
are continuing to operate under the assumption of FD-shaped distributions as
discussed previously (we will relax this constraint in Sec.\ \ref{sec:trans}),
implying we can write \snu as
\beq
  \snu=\frac{2\pi^2}{45}\frac{7}{8}6T_\nu^3,
\eeq
where the factor of 6 comes from 3 flavors of neutrinos, and 3 flavors of
antineutrinos all at the same temperature parameter.  We have used the symbol
$T_\nu$ to denote the neutrino temperature parameter
\beq
  T_\nu(a) = \begin{cases}
  T(a) & a>\adec\\
  \tcm(a) & a<\adec
  \end{cases}.
\eeq
In practice, we will refrain from using the symbol $T_\nu$ and instead use
either $T(a)$ or $\tcm(a)$ to denote the energy scale at a particular epoch.

\begin{figure}
  \includegraphics[width=\columnwidth]{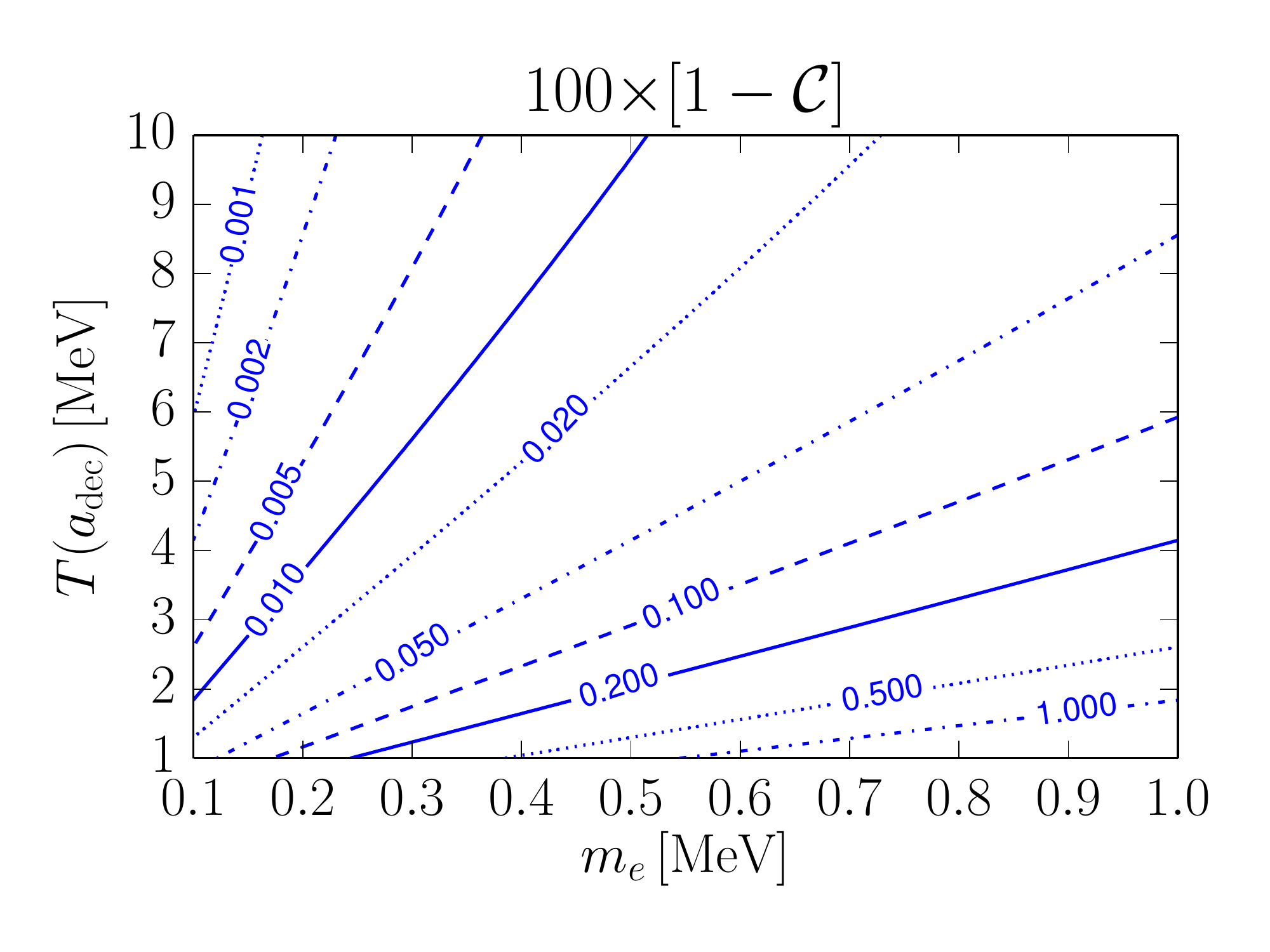}
  \caption{\label{fig:tdec_vs_xm_contour_sencorr} 
  Contours of $100\times[1-\mathcal{C}]$ plotted in the \tdec versus $m_e$
  plane.  $\mathcal{C}$ is the correction to the baryon number density
  in Eq.\ \eqref{eq:bn5}.  The contours are calculated using the full \burst code
  and agree with the analytic estimate in Eq.\ \eqref{eq:c_corr}}
\end{figure}

We begin executing \burst at a temperature higher than \tdec.  This is for
computational reasons only: our code must initialize a time step before we
consider the cosmological epochs relevant to whatever physics we wish to study.
As a corollary, we normalize the baryon number density at the starting
temperature $\tstart=T(a_i)$ in Eq.\ \eqref{eq:bn4}.  At \tstart, the total
entropy in the universe is
\beq\label{eq:stot1}
  \stot(a_s) = \frac{1}{n_b(a_s)}\frac{2\pi^2}{45}\left\{2+
  \frac{7}{8}[2+2]\left[1-\frac{15}{14\pi^2}x^2(a_s)\right] +
  \frac{7}{8}6\right\}\tstart^3,
\eeq
where $\xstart\equiv m_e/\tstart$.  The total entropy at the current epoch
(well after electrons and positrons have annihilated) is
\beq\label{eq:stot2}
  \stot(a_0) = \frac{1}{n_b(a_0)}\frac{2\pi^2}{45}\left\{2+
  \frac{7}{8}6\left[\frac{\tcm}{T}\right]^3_{\rm f.o.}\right\}T_0^3.
\eeq
If we use Eq.\ \eqref{eq:tratio4} and equate Eqs.\ \eqref{eq:stot1} to
\eqref{eq:stot2}, we find
\begin{align}
  n_b(a_{s}) &= n_b(a_0)\frac{11}{4}
  \left[1 - \frac{15}{43\pi^2}x^2(a_s) -
  \frac{315}{946\pi^2}x^2(\adec)\right]
  \left[\frac{\tstart}{T(a_0)}\right]^3,\\
  \implies\mathcal{C} &= 
  1 - \frac{15}{43\pi^2}x^2(a_s)
  - \frac{315}{946\pi^2}x^2(\adec).\label{eq:c_corr}
\end{align}
Figure \ref{fig:tdec_vs_xm_contour_sencorr} shows contours of constant
$100\times[1-\mathcal{C}]$ in the $T(\adec)$ versus $m_e$ parameter space.  The
contours were calculated using \burst, and agree exceedingly well with the
prediction of Eq.\ \eqref{eq:c_corr}.  If we only used the correction from
\tcmplfo, and neglected the contribution from $x(a_s)$, the contours from the
calculation would have diverged from Eq.\ \eqref{eq:c_corr} at large $m_e$.

\section{Finite Temperature QED corrections}\label{sec:qed}

\begin{figure}
  \begin{center}
    \includegraphics[width=\columnwidth]{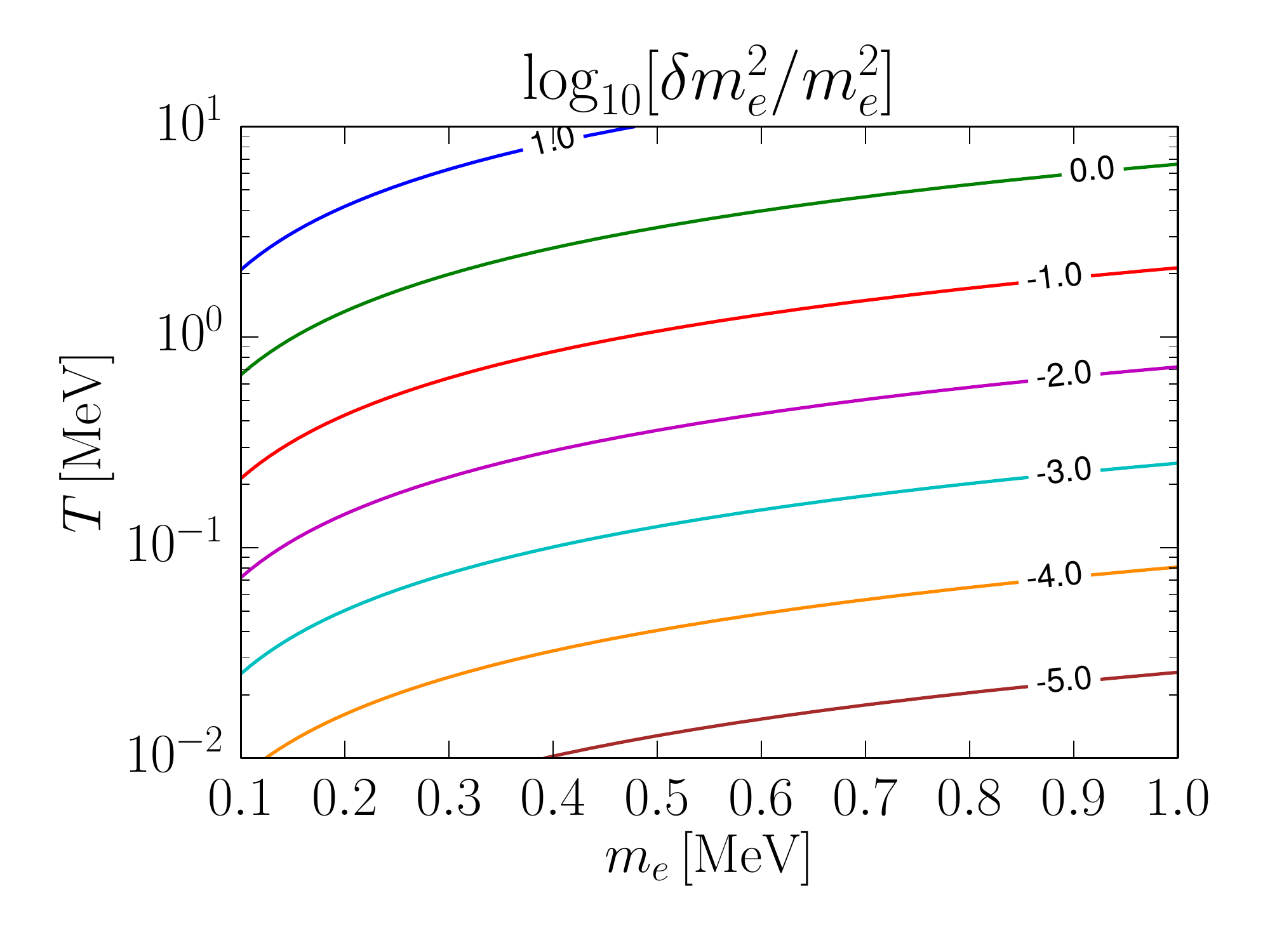}
  \end{center}
  \caption{\label{fig:t_vs_xm_dme2me} Contours of $\log_{10}[\delta
  m_e^2/m_e^2]$ plotted in the $T$ versus $m_e$ plane.  This figure is
  a plot of Eq.\ \eqref{eq:dme1} with $p=0$.}
\end{figure}

Thus far, we have taken the gas of photons, electrons, and positrons to behave
like an ideal gas, with energy density for fermions from Eq.\ \eqref{eq:rho_def} and
pressure for fermions and bosons given by 
\beq\label{eq:p_def}
  P_j^{(0)} = \frac{g_j}{2\pi^2}\int_0^{\infty}
  dp \frac{p^4}{3E(p)}\frac{1}{e^{E(p)/T}\pm1},
\eeq
where $E(p)=\sqrt{p^2 + m_j^2}$ for species $j$.  The ``+'' sign in the
occupation number refers to fermions (electrons or positrons), whereas the
``$-$'' sign refers to bosons (photons).  We have ignored the chemical potential
for electrons and positrons.  The superscript $(0)$ on the pressure symbol
denotes that Eq.\ \eqref{eq:p_def} is the pressure for the ideal gas.  In the
early universe, charge screening and self-interaction energies will change the
pressure quantity.  Reference \cite{1994PhRvD..49..611H}  gives the change to
the pressure as
\begin{align}
  P_j &= P_j^{(0)} - P_j^{{\rm (int)}}\\
  P_j^{{\rm (int)}} &= \frac{1}{4\pi^2}\int_0^{\infty}dp
  \frac{p^2}{E(p)}\frac{\delta m_j^2(p,T)}{e^{E(p)/T}\pm1},\label{eq:pint}
\end{align}
which introduces the shift in the square of the particle mass, $\delta
m_i^2(p,T)$, as a function of $p$ and $T$.  Reference
\cite{1994PhRvD..49..611H}  calculates the shift from QED self-interactions for
electrons or positrons as
\begin{align}
  \delta m_e^2(p,T) = \frac{2\pi\alpha T^2}{3}
  &+\frac{4\alpha}{\pi}\int_0^{\infty}dk\frac{k^2}{E(k)}
  \frac{1}{e^{E(k)/T}+1}\nonumber\\
  &- \frac{2m_e^2\alpha}{\pi p}\int_0^{\infty}dk
  \frac{k}{E(k)}\log\left|\frac{p+k}{p-k}\right|
  \frac{1}{e^{E(k)/T}+1}.\label{eq:dme1}
\end{align}
$m_e$ in the above expression is still the vacuum mass and
$E(k)=\sqrt{k^2+m_e^2}$.  $\alpha\simeq 1/137$ is the fine structure constant.
As we do not include the electron chemical potential, the shift in the square
of the electron and positron masses are identical.  Initially, we will ignore
the third term in Eq.\ \eqref{eq:dme1}; this is tantamount to $\delta
m_e^2(p,T)\rightarrow\delta m_e^2(T)$. Figure \ref{fig:t_vs_xm_dme2me} shows
how Eq.\ \eqref{eq:dme1} changes in the $T$ versus $m_e$ parameter space.  Note
that $\delta m_e^2$ becomes larger with increasing temperature.  $\delta m_e^2$
is equal to the vacuum value of $m_e^2$ at a temperature of a few MeV.  We have
shown that the ratio \tcmplfo is sensitive to the thermodynamics in this
temperature range.  For high precision calculations of neutrino energy density,
an accurate description of this epoch is imperative.

Photons in medium are plasmons with effective mass squared given by
\cite{1997PhRvD..56.5123F}
\beq\label{eq:dmp}
  \delta m_\gamma^2(T) =
  \frac{8\alpha}{\pi}\int_0^{\infty}dk\frac{k^2}{E(k)}\frac{1}{e^{E(k)/T}+1}.
\eeq
Note that the shift in mass from zero is solely a function of temperature.

With the relevant in-medium particle masses inferred from these shifts, the
entropy per baryon at \adec will be slightly different, which will lead to a
change in \tcmplfo.  We can analytically estimate the contribution of $P^{\rm
(int)}$ to the total pressure if we assume $E(p)\simeq p$ in Eq.\
\eqref{eq:pint} and $E(k)\simeq k$ in Eqs.\ \eqref{eq:dme1} and \eqref{eq:dmp}.
After applying the approximations, the shifts in the masses are
\begin{align}
  \delta m_e^2(T) &= \pi\alpha T^2,\\
  \delta m_\gamma^2(T) &= \frac{2\pi\alpha T^2}{3},
\end{align}
and the interacting pressures are
\begin{align}
  P^{\rm (int)}_e &= \frac{\pi\alpha T^4}{48},\\
  P^{\rm (int)}_\gamma &= \frac{\pi\alpha T^4}{36},
\end{align}
where $P^{\rm (int)}_e$ is the interacting pressure for either electrons or
positrons.  Now that we have the expressions for the various $P^{\rm (int)}_i$,
we need an expression for the energy density.  To calculate the total energy
density, we minimize the Gibbs free energy to find \cite{Mangano:3.040}
\beq
  \rho = -P + T\frac{dP}{dT}.
\eeq
If we use the above approximations, the derivatives are trivial and the
interacting energy densities become
\begin{align}
  \rho^{\rm (int)}_e &= \frac{\pi\alpha T^4}{16},\\
  \rho^{\rm (int)}_\gamma &= \frac{\pi\alpha T^4}{12}.
\end{align}
The interacting pressure and energy density components give us the conserved
entropy per baryon in terms of $\spl^{(0)}$
\beq\label{eq:spl_qed}
  \spl = \spl^{(0)}\left[1 - \frac{25\alpha}{22\pi}\right].
\eeq
To calculate a new \tcmplfo ratio, we use Eq.\ \eqref{eq:tcmpl_z} and the
correction from Eq.\ \eqref{eq:spl_qed} to find
\beq\label{eq:tcmpl_qed}
  \left[\frac{\tcm}{T}\right]_{\rm f.o.} 
  = \left[\frac{4}{11}\right]^{1/3}
  \left[1+\frac{5}{22\pi^2}x^2(\adec)
  +\frac{25\alpha}{66\pi}\right].
\eeq
The QED contribution to the change in \tcmplfo is identical to that of Eq.\
(41) in Ref.\ \cite{1999PhRvD..59j3502L}.

In Sec.\ \ref{sec:notrans}, we estimated the change in the correction to the
baryon density from the electron rest mass.  As an alternative to baryon
density, we can use the radiation energy density parameterized by \neff
\beq\label{eq:neff1}
  \rho_{\rm rad}(a) = \left\{2+\frac{7}{4}
  \left[\frac{4}{11}\right]^{4/3}\neff\right\}\frac{\pi^2}{30}T^4(a).
\eeq
$\rho_{\rm rad}$ is the radiation energy density, with photon and neutrino
components.  We could use Eq.\ \eqref{eq:neff1} at any value of the scale
factor and monitor how \neff evolves as the neutrinos decouple from the
electromagnetic plasma [e.g., see Fig.\ (5) of Ref.\ \cite{transport_paper}].
Generally, \neff is considered a constant and Eq.\ \eqref{eq:neff1} is used
once \tcmplfo nears its asymptotic value.  In this paper, we will adopt the
traditional approach and consider how \tcmplfo and \neff change in the
$T(\adec)$ versus $m_e$ parameter space.  Equation (31) of Ref.\
\cite{transport_paper} provides an expression for \neff which we can solve when
the ratio of \tcmplfo differs from \felev
\beq\label{eq:neff2}
  \neff = \{1+\delta\tcmplfo\}^4\times
  [3+\delta\rho_{\nu_e}+\delta\rho_{\nu_\mu}+\delta\rho_{\nu_\tau}],
\eeq
where $\delta\rho_{\nu_i}$ is the relative change in the neutrino energy
density of species $i$ at freeze-out and we have assumed that neutrinos and
antineutrinos of any flavor change identically. If neutrinos were to preserve
FD-shaped distributions, $\delta\rho_{\nu_i}=0$. Therefore, the change in
\neff stemming from nonzero rest mass and QED corrections is
\begin{align}
  \Delta\neff\equiv\neff-3 &=3\times\{\delta\tcmplfo\}^4\\
  &\simeq\frac{30}{11\pi^2}x^2(\adec) + \frac{50\alpha}{11\pi}.
  \label{eq:dneff}
\end{align}

\begin{figure}
  \begin{center}
    \includegraphics[width=\columnwidth]{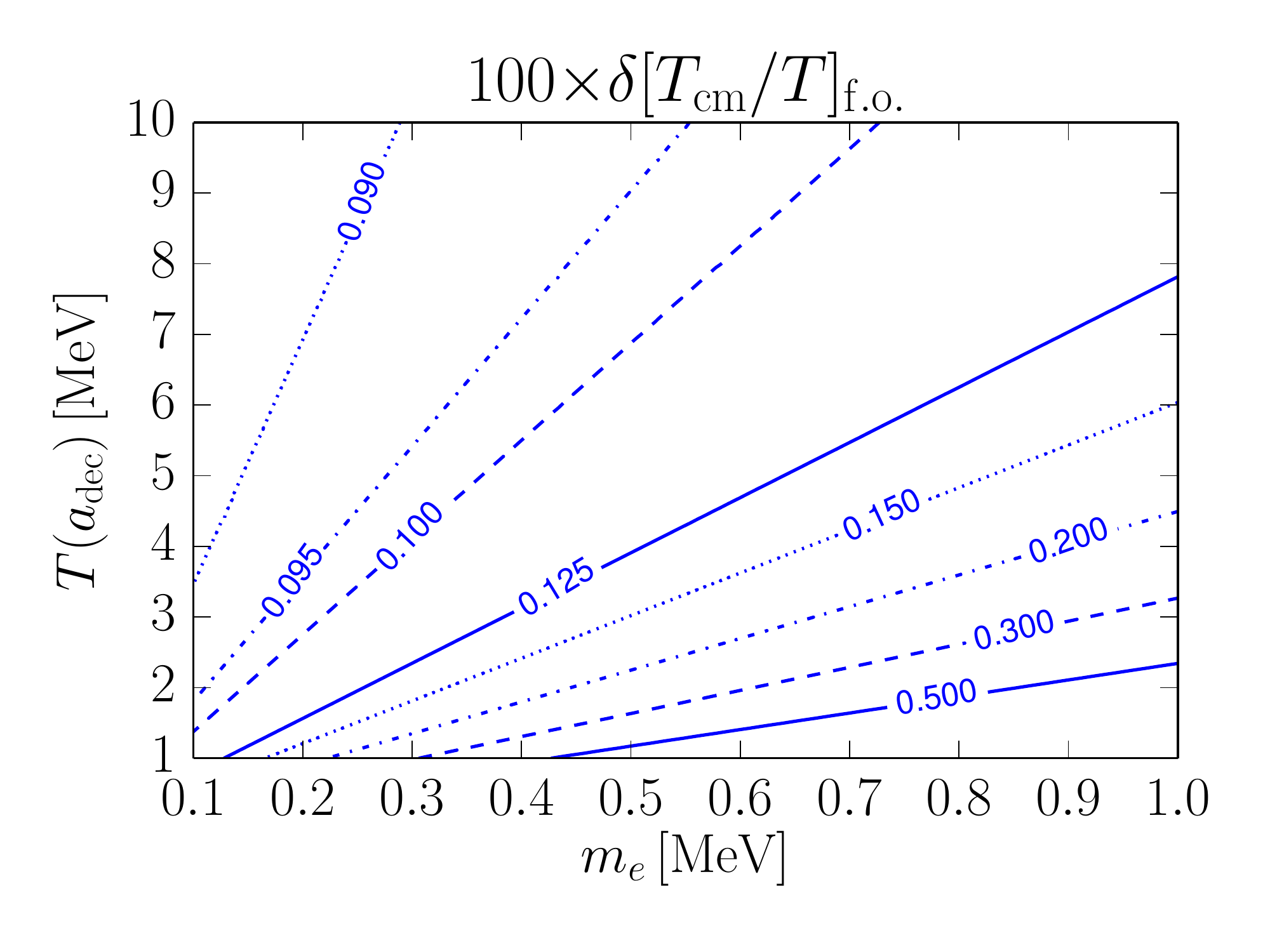}
    \includegraphics[width=\columnwidth]{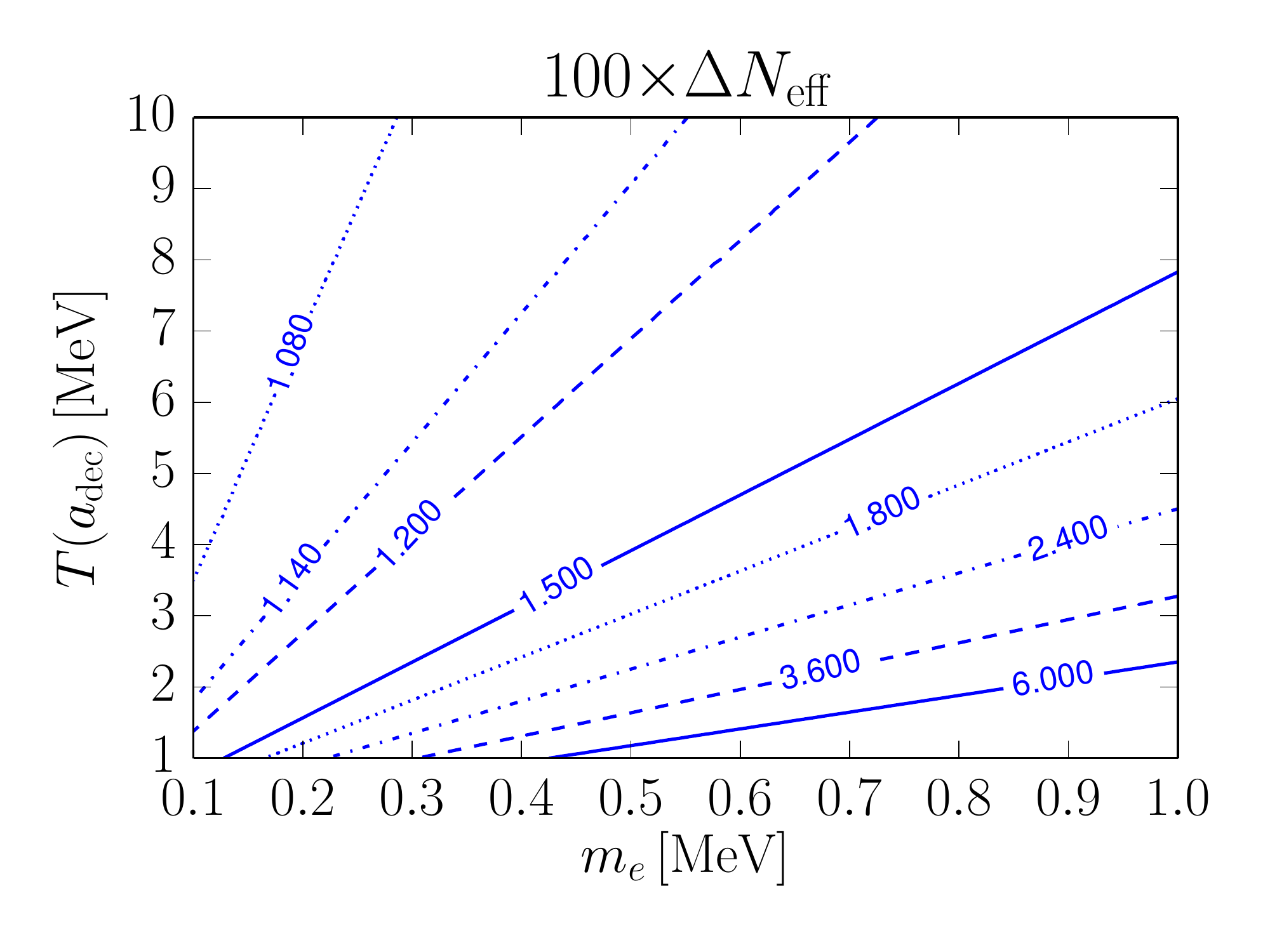}
  \end{center}
  \caption{\label{fig:tcmpl_neff_qed} Plots of freeze-out quantities with
  finite-temperature QED effects.  (Top) Contours of $100\times\delta\tcmplfo$ in
  the $T(\adec)$ versus $m_e$ plane.  The contours are calculated using \burst,
  however, they agree with the analytic estimate in Eq.\ \eqref{eq:tcmpl_qed}.
  (Bottom) $100\times\Delta\neff$ in the $T(\adec)$ versus $m_e$ plane.  The
  contours are calculated using \burst, however, they agree with the analytic
  estimate in Eq.\ \eqref{eq:dneff}.}
\end{figure}

Figure \ref{fig:tcmpl_neff_qed} shows the changes in \tcmplfo and \neff as
calculated in \burst.  The top panel gives $100\times\tcmplfo$ in the
$T(\adec)$ versus $m_e$ parameter space.  The contours agree well with the
analytic estimate of Eq.\ \eqref{eq:tcmpl_qed}.  As compared to Fig.\
\ref{fig:tdec_vs_xm_contour_tcmpl}, the contours do not cover as large a range
of values.  For small $m_e$, the QED correction dominates and the contours
become spaced further apart, never reaching the $10^{-4}$ level as they do in
Fig.\ \ref{fig:tdec_vs_xm_contour_tcmpl}.  Conversely, for large $m_e$, the
contribution from the vacuum rest mass becomes dominant and the contours begin
to look identical to those of \ref{fig:tdec_vs_xm_contour_tcmpl}.  The bottom
panel shows how $100\times\Delta\neff$ changes in the same parameter space.  If
we use the small-value approximation for a power function, i.e.,
$[1+y]^n\simeq1+ny$ for $y<<1$, Eq.\ \eqref{eq:neff2} shows
$\Delta\neff\simeq12\delta\tcmplfo$.  We chose the contours of the bottom plot
of Fig.\ \ref{fig:tcmpl_neff_qed} to be 12 times the value of the contours in
the top plot.  Clearly, corresponding contours appear in the identical parts of
the parameter space.

In this section we ignored how $\delta m_e^2$ depends on momentum $p$ in Eq.\
\eqref{eq:dme1} when deriving the correction to \tcmplfo.  We have done the
full calculation with the momentum dependence included and have found a
relative change in $\Delta\neff$ of less than one part in $10^{3}$.  A change
this small would only be discernible in Fig.\ \ref{fig:tcmpl_neff_qed} for
small $m_e$.

\section{Full Neutrino Transport and Nonequilibrium Energy
Distributions}\label{sec:trans}

Here we consider the timelike heat flow engendered by out-of-equilibrium
neutrino scattering on electrons and positrons and the associated alterations
in comoving entropy conservation, \neff, the neutron-to-proton ratio, and
primordial nucleosynthesis.  As the universe expands and the
temperature decreases, the equilibrium between photons and $e^\pm$ pairs
shifts to fewer of the later, ultimately increasing the product of scale factor
and plasma temperature.  However, while the annihilation of electrons and
positrons proceeds, neutrinos are going from completely-thermally coupled
particles to freely-falling decoupled particles.  Electrons and positrons are
the principal scattering targets that facilitate energy exchange between the
decoupling neutrinos and the plasma.  These scattering processes occur out of
equilibrium, thereby generating a timelike heat flow, i.e., transfer of
entropy, from the plasma into the decoupling neutrino component.

The heat transfer between $e^\pm$ pairs and neutrinos results in a
decrease of \spl and an increase in \snu.  Overall, the total entropy per
baryon of the universe, $\snu+\spl$, increases [see Fig.\ (9) of Ref.\
\cite{transport_paper}].  The expressions we derived in sections
\ref{sec:notrans} and \ref{sec:qed} all assumed comoving entropy conservation
in either the special case of the plasma or the general case of the universe in
total.  In either case, those expressions are not valid during the weak
decoupling process.

As the neutrinos do not maintain FD equilibrium in this case, we must calculate
the effective entropy per baryon in the neutrino seas using a
energy-distribution based definition of entropy density for fermions
\beq
  \snu(a)=-\frac{1}{n_b(a)}\frac{\tcm^3(a)}{2\pi^2}\sum\limits_{j=1}^6
  \int_0^\infty d\epsilon\,\epsilon^2\{f_j\ln f_j + [1-f_j]\ln[1-f_j]\},
\eeq
where we have implicitly assumed that $a>\adec$.  $\epsilon$ is a dummy
variable such that $\epsilon=E/\tcm$ and $f_j$ is the occupation number as a
function of $\epsilon$ for the six individual neutrino species $j$.  In
addition, the $f_j$ evolve and are functions of $a$.  The Quantum Kinetic
Equations (QKEs) dictate the evolution of the neutrino occupation numbers
\cite{Blaschke_Cirigliano_2016}
\beq
  \frac{\partial \hat{F}}{\partial t} = -i[H,\hat{F}] + \hat{C},
\eeq
where $i=\sqrt{-1}$ and we have taken Eq.\ (24) of
\cite{Blaschke_Cirigliano_2016} and simplified to a homogeneous and isotropic
geometry.  We have departed from our convention of using $a$ as the independent
variable in order to be consistent with QKE literature (see Refs.\
\cite{Blaschke_Cirigliano_2016,1991NuPhB.349..743B,AkhmedovBerezhiani,
1993APh.....1..165R,2005PhRvD..71i3004S,2007JPhG...34...47B,2013PhRvD..87k3010V,
2013PrPNP..71..162B,Gouvea,VFC:QKE,2014PhRvD..90l5040S,2015PhRvD..91l5020K,
2015PhLB..747...27C, 2015PhyS...90h8008K,2015IJMPE..2441009V,
2016NuPhB.908..366C,2016JCAP...12..019S} for a discussion on QKEs).  $\hat{F}$
is a $6\times6$ generalized density matrix where the occupation numbers of the
3 neutrinos and 3 antineutrinos fall along the diagonal of the matrix.  The
coherent evolution of \fhat is given by the commutator of a Hamiltonian-like
potential with \fhat.  We will ignore this term and focus on the term which can
affect the entropy\footnote{ Neutrino oscillations, via the Hamiltonian-like
potential, can indeed induce changes in the entropy.  We ignore those
contributions.  See Eq.\ (5.12) in Ref.\ \cite{1993NuPhB.406..423S} for
details.}, mainly the collision term $\hat{C}$, which encodes incoherent
neutrino scattering collisions.  As we ignore the coherent evolution, the
off-diagonal elements of \fhat remain zero thereby lessening the need for a
full QKE treatment.  We employ a collision integral, $C_j$, originally derived
in Refs.\ \cite{1995PhRvD..52.1764H,Dolgov:1997ne} and modified in Ref.\
\cite{transport_paper}
\beq
  \frac{df_j(\epsilon)}{dt}= C_{j}(\{f_k\}),
\eeq
where the collision integral for $f_j(\epsilon)$ is a functional of the entire
set of occupation numbers $\{f_k\}$.  The change in the energy density of the
neutrino component is
\begin{align}
  \frac{d\rho_\nu}{dt} &= \frac{\tcm^4}{2\pi^2}\sum\limits_j
  \int_0^{\infty}d\epsilon\,\epsilon^3\frac{df_j(\epsilon)}{dt}\\
  &= -\frac{d\rho_e}{dt}.\label{eq:drhoedt}
\end{align}
Here $d\rho_e/dt$ represents the instantaneous decrement in the energy density
residing in the electron and positron components.  Neutrino-electron
scattering, for example, might result in a higher energy neutrino and a lower
energy electron, leaving a nonthermal energy distribution for the electrons.
We assume that thermal and chemical equilibrium in the electromagnetic plasma
is instantaneously reattained.  To properly follow the evolution of the
temperature, we add this change in energy density from Eq.\ \eqref{eq:drhoedt}
into the plasma temperature derivative \cite{letsgoeu2}
\beq
  \frac{dT}{dt} \simeq -3H\,
  \frac{\rho + P - \dfrac{1}{3H}\dfrac{d\rho_e}{dt}}
  {\dfrac{d\rho}{dT}},\label{eq:dtempdt}
\eeq
where $\rho$ and $P$ are the energy density and pressure, respectively, of the
electromagnetic plasma.  We note that Eq.\ \eqref{eq:dtempdt} also contains
terms for the baryon components.  We have ignored these terms when writing
Eq.\ \eqref{eq:dtempdt} for ease in notation, but we include them in the actual
calculation [see Eq.\ (D.28) in Ref.\ \cite{letsgoeu2}].  In summary, the
change in the plasma energy density from neutrino scattering decreases the
temperature and hence raises the ratio \tcmplfo.  

\begin{table*}
  \begin{center}
  \begin{tabular}{c c c c c}
    \hline
    $m_e$ [MeV] & \tcmplfo & $100\times[1-\mathcal{C}]$ & $\Delta\neff$ & $Y_P$\\
    \midrule[1.0pt]
    0.1 & 0.7144 & 0.2808 & 0.0120 & 0.3668\\
    0.25 & 0.7146 & 0.3655 & 0.0190 & 0.3313\\
    0.511 & 0.7154 & 0.6610 & 0.0442 & 0.2479\\
    0.75 & 0.7164 & 1.069 & 0.0793 & 0.1510\\
    1.0 & 0.7178 & 16.06 & 0.1265 & 0.0441\\
    \hline
  \end{tabular}
  \end{center}
  \caption{\label{tab:trans_runs}Various quantities of interest for each $m_e$
  run with neutrino transport.  Column 2 gives \tcmplfo.  For reference:
  $[4/11]^{1/3}=0.7138$.  Column 3 gives the correction factor for the baryon
  number density in Eq.\ \eqref{eq:bn5}.  Column 4 gives the change in \neff from
  Eq.\ \eqref{eq:neff2}.  Column 5 is the primordial mass fraction of \heiv.
  $\tin=10\,{\rm MeV}$ for all runs.
  }
\end{table*}

Once we have the derivative for the plasma temperature, we can evolve through
the BBN epoch to determine \tcmplfo.  Table \ref{tab:trans_runs} gives
quantities related to the topics previously discussed for five different
assumed values of $m_e$.  In the third row, $m_e=0.511\,{\rm MeV}$ corresponds
to the true vacuum value.  The calculations include the integration of the full
Boltzmann neutrino transport network of Ref.\ \cite{transport_paper} and the
finite-temperature QED corrections, including the momentum-dependent term in
Eq.\ \eqref{eq:dme1}.  We initiate the Boltzmann neutrino transport calculation
at a temperature sufficiently high such that the neutrino component is in
equilibrium.  We take this epoch to be the same as the one designated in Eq.\
\eqref{eq:tcm}.  For all values of $m_e$, we take $\tin=10\,{\rm MeV}$.  We
give the ratio \tcmplfo in column 2, where we calculate the evolution of $T$
with the $d\rho_e/dt$ term in Eq.\ \eqref{eq:dtempdt}.  Column 3 relates to the
correction factor for the baryon number density, defined in Eq.\
\eqref{eq:bn5}, and column 4 gives the calculation of $\Delta\neff$.  Here,
$\Delta\neff$ includes not only the finite electron mass and QED corrections
discussed above, but also the scattering-induced, nonthermal neutrino energy
distribution functions \footnote{The third row in the table shows $\neff=3.044$
in the standard case.  This is different than the previous results of Refs.\
\cite{transport_paper} and \cite{lep_trans}.  This difference stems from an
alternate implementation of the finite-temperature QED effects. The new
implementation is an attempt to conform to Ref.\ \cite{Mangano:3.040}.
However, our implementation handles the derivative of temperature with respect
to time differently than is done in Ref.\ \cite{Mangano:3.040} [see Eq.\
\eqref{eq:dtempdt} and Eq.\ (17) in that work].  Note that our calculation
neglects all neutrino flavor oscillations.  As a consequence of these
considerations, we would not expect to agree with Refs.\ \cite{neff:3.046}
($\neff=3.046$) and \cite{2016JCAP...07..051D} ($\neff=3.045$) to the $10^{-3}$
level of precision. However, we note that our value of \neff does agree with
Ref.\ \cite{2015NuPhB.890..481B} to the $10^{-3}$ level.}.  All three of the
quantities in columns 2 -- 4 involve a set of three corrections: nonzero
electron mass; finite-temperature QED effects on the plasma equation of state;
and entropy flow between the plasma and neutrino seas.  The contribution of the
entropy flow to $\delta\tcmplfo$, $1-\mathcal{C}$, and $\Delta\neff$ is on
order a factor of 3 - 50 times larger than the contribution from the other two
effects, i.e., nonzero electron mass and finite-temperature QED effects.
Appendix B of Ref.\ \cite{1982PhRvD..26.2694D} gives an analytic estimate for
how the entropy flow changes \tcmplfo.  Finally, column 5 gives the primordial
abundance of \heiv, $Y_P$, from nucleosynthesis.

\begin{figure}
  \begin{center}
    \includegraphics[width=\columnwidth]{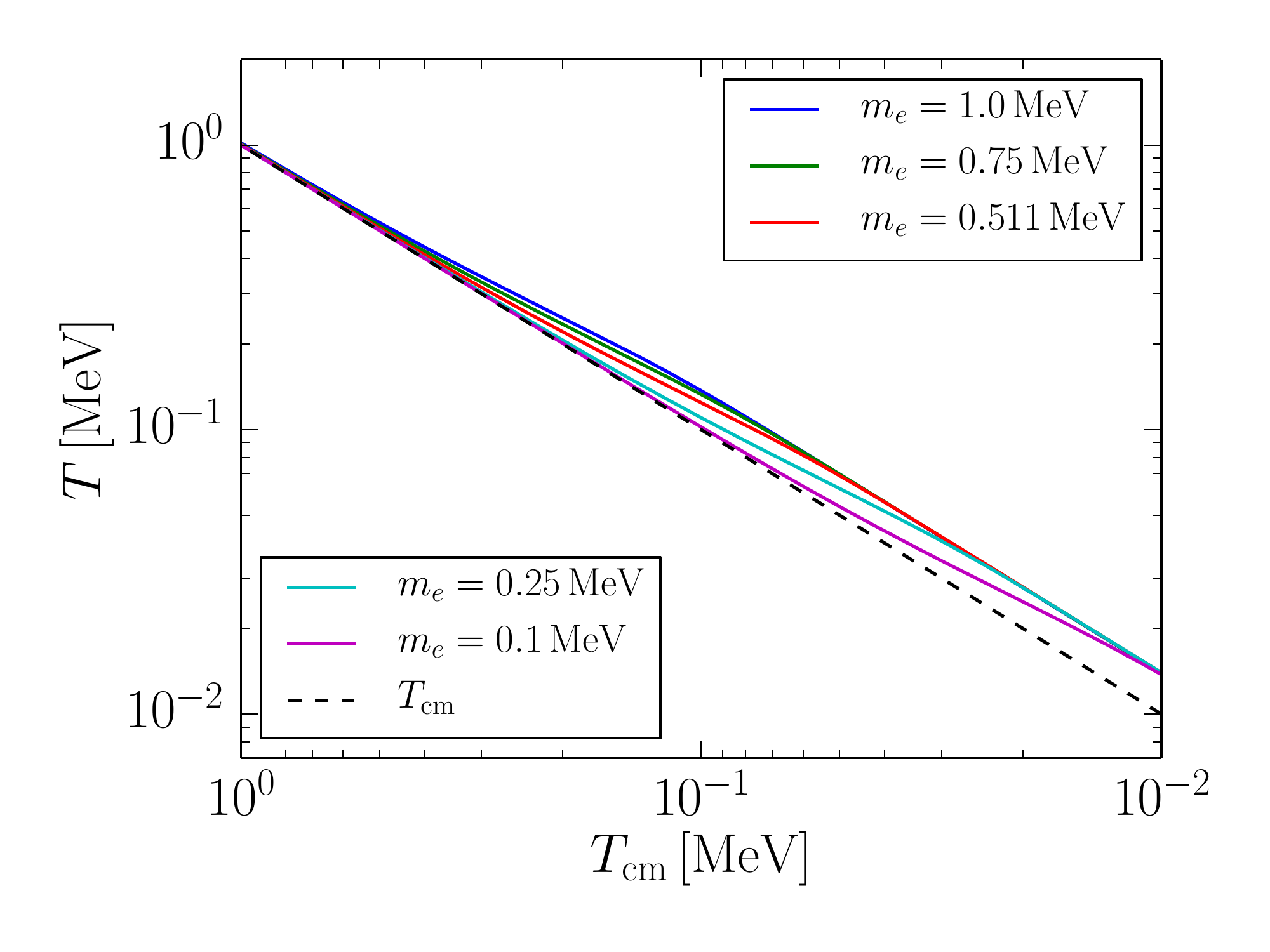}
    \includegraphics[width=\columnwidth]{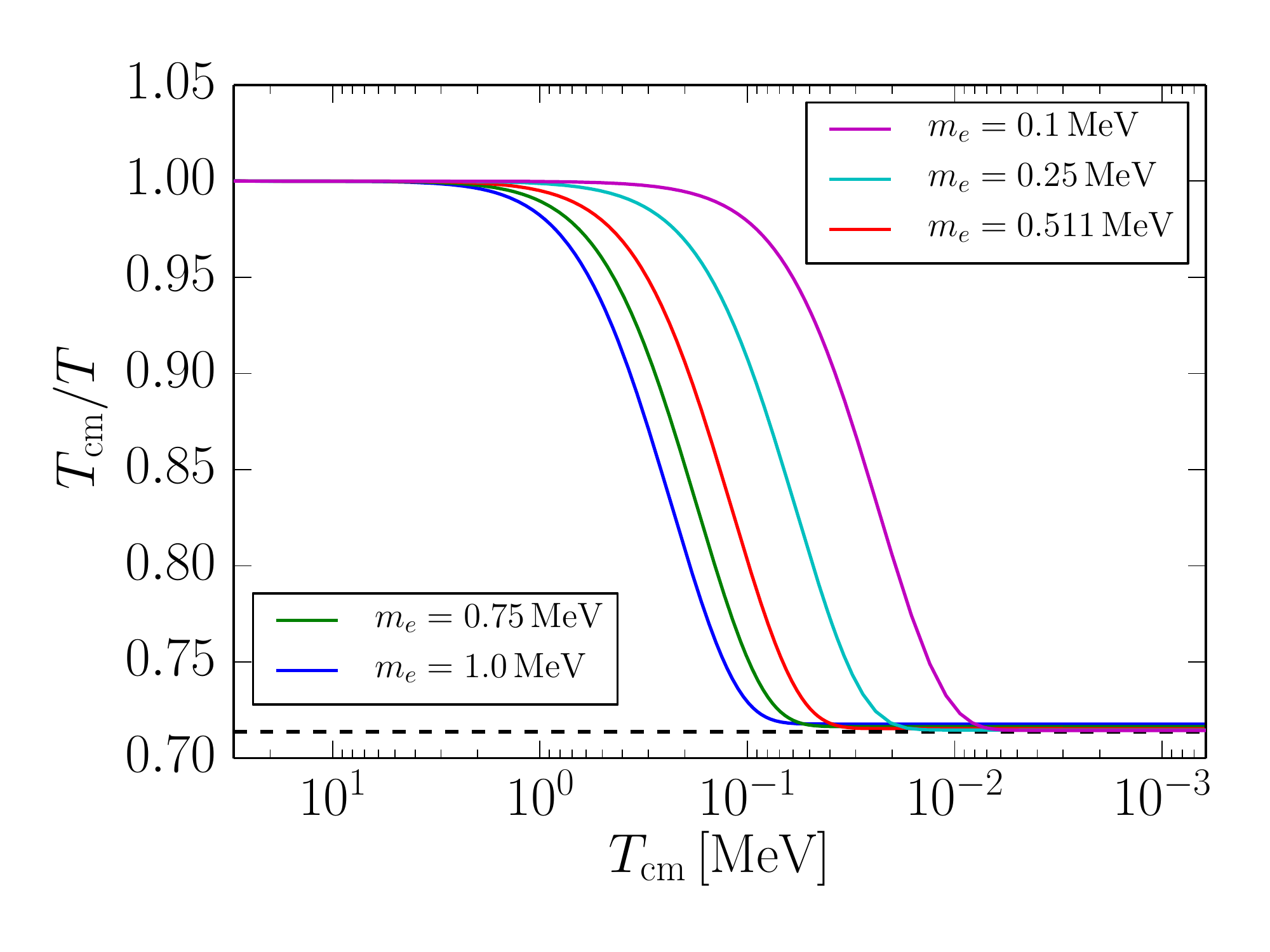}
  \end{center}
  \caption{\label{fig:temps_vs_tcm} Plots of temperature quantities with
  finite-temperature QED effects and neutrino energy transport.  (Top) Plasma
  temperature as a function of \tcm.  The dashed black line is \tcm plotted
  against \tcm.  (Bottom) Ratio of comoving temperature parameter to plasma
  temperature as a function of \tcm.  The dashed black line is the ratio
  $\felev=0.7138$.}
\end{figure}

Figure \ref{fig:temps_vs_tcm} shows quantities involving the plasma temperature
as a function of \tcm plotted for various values of $m_e$.  The top panel shows
how $T$ cools as the universe expands, while the bottom panel shows how the
neutrinos cool relative to the plasma via the ratio \tcmpl.  The horizontal
axis in both plots is \tcm.  The five colored lines show the evolution of
either $T$ or \tcmpl for various values of $m_e$.  At the level of precision in
the top panel, it appears that all five lines for $T$ converge at the end of
the epoch of $e^\pm$ annihilation, implying identical values of \tcmplfo.  The
second column of Table \ref{tab:trans_runs} shows that there are slight
differences due to the effects discussed previously.  This is more apparent at
the level of precision in the bottom panel, where we plot the ratio \tcmpl
versus \tcm.  As the mass increases, the ratio at freeze-out also increases.
Although the mass changes \tcmplfo, both panels vividly show how the mass
changes the {\it location} of the epoch of $e^\pm$ annihilation.

Note that Fig.\ \ref{fig:temps_vs_tcm}, consistent with Fig.\
\ref{fig:nep_vs_tcm}, shows the large number of $e^\pm$ pairs in
equilibrium even at temperatures well below the electron rest mass.
Additionally, Fig.\ \ref{fig:temps_vs_tcm} shows how the large
$e^\pm$ density, a consequence of the high entropy, facilitates
transfer of entropy from the electromagnetic plasma into the decoupling
neutrino component.  As we assume FD equilibrium for the $e^\pm$ occupation
numbers, a larger mass will precipitate an earlier epoch/higher temperature
when the $e^\pm$ pairs disappear.  This leads to a different phasing of \tcm
and $T$ clearly shown in the bottom panel of Fig.\ \ref{fig:temps_vs_tcm}.

\begin{figure}
  \includegraphics[width=\columnwidth]{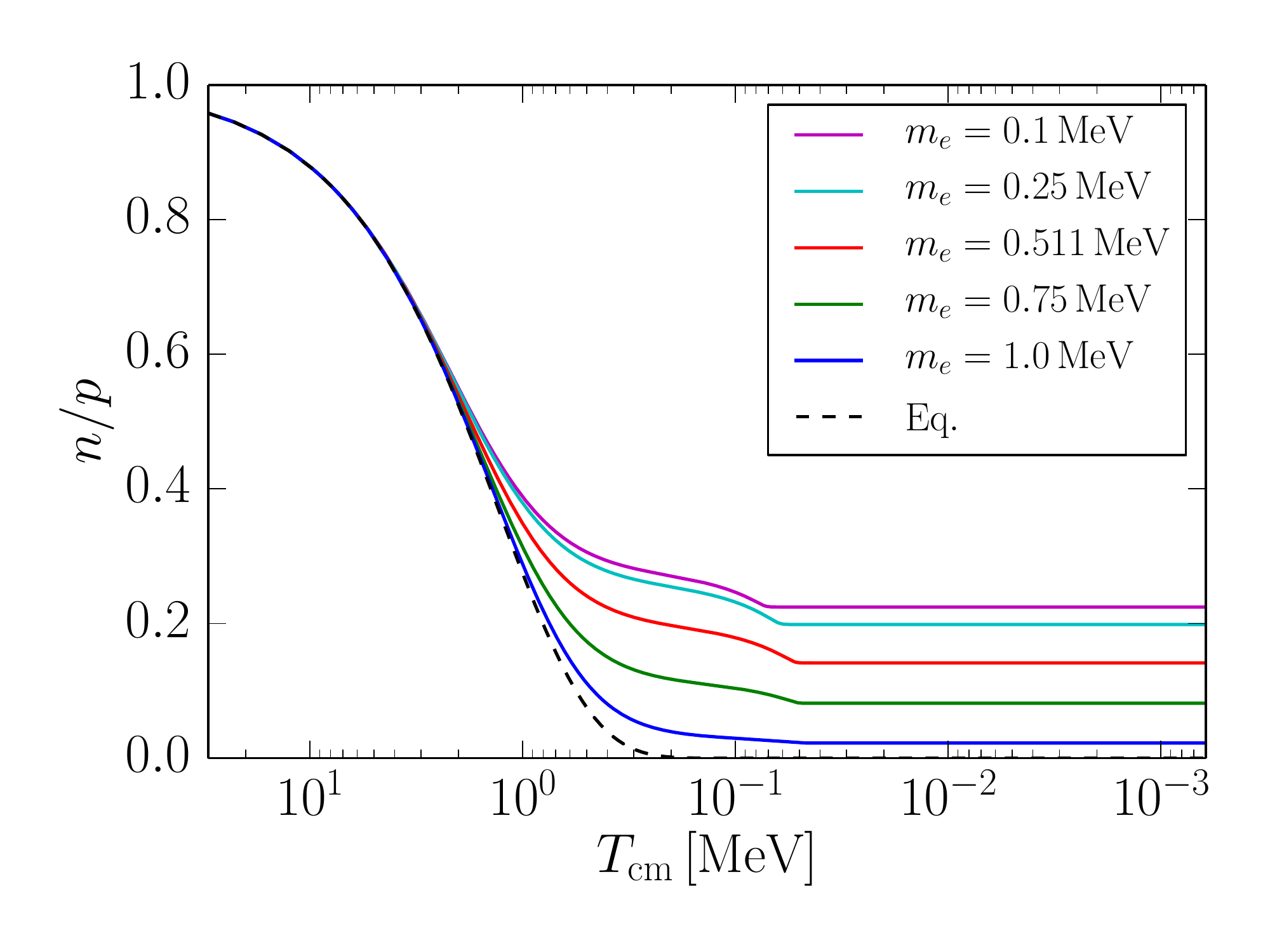}
  \caption{\label{fig:np_vs_tcm} Neutron-to-proton ratio as a function of \tcm.
  The dashed black line is the equilibrium evolution of \np in the case
  $m_e=1.0\,{\rm MeV}$ [see Eq.\ \eqref{eq:np_eq}].}
\end{figure}

The phasing of \tcm and $T$ is important as both energy scales are inputs into
the weak-interaction rates which interconvert neutrons and protons.  Before we
present results related to nucleosynthesis, we note that only the assumed
electron vacuum mass is used in the calculations of the weak-interaction rates;
we do not use the finite-temperature QED modifications for the weak-interaction
rates \cite{FFN:I,FFN:III,FFN:IV}.  Figure \ref{fig:np_vs_tcm} shows the
neutron-to-proton ratio (denoted $n/p$) as a function of \tcm for the various
masses.  As the mass increases, the epoch of $e^\pm$ annihilation moves earlier
in time, higher in \tcm.  This leads to reduced efficiency in
scattering-induced transfer of entropy from the electromagnetic plasma into the
neutrino seas.  In turn, this effect leads to generally higher plasma
temperatures at a given epoch \tcm.  Conversely, we could say that the earlier
epoch of $e^\pm$ annihilation leads to smaller \tcm at earlier times, but this
is not the correct way to think about this problem/effect.  The product of \tcm
and scale factor is a comoving invariant, i.e., $\tcm a=$constant.  When we
compare the neutron to proton interconversion rates [see Eqs.\ (19) -- (24) in
Ref.\ \cite{WFO_approx}] at equivalent times/scale-factors, we are comparing at
the same \tcm, so larger $T$ or smaller \tcm are not equivalent statements.  If
the temperature is larger, there is an enhancement in the two charged-lepton
capture rates: $e^+ + n\rightarrow p + \bnue$ and $e^- + p\rightarrow n +
\nue$.  (Note that $m_e$ is also larger which would suppress the charged-lepton
capture rates; however, the increase in temperature is more important at higher
kinetic energies).  Both of these rates are enhanced and keep $n/p$ in
equilibrium to lower temperatures
\beq\label{eq:np_eq}
  (n/p)^{\rm (eq)} = e^{-\delta m_{np}/T},
\eeq
where $\delta m_{np}\simeq 1.3\,{\rm MeV}$ is the mass difference between a
neutron and a proton, and we have neglected electron and neutrino degeneracies.
The black dashed line in Fig.\ \ref{fig:np_vs_tcm} gives $(n/p)^{\rm (eq)}$ for
the case where we evaluate the evolution of the temperature $T$ with assumed
electron mass $m_e=1.0\,{\rm MeV}$.  The curve for $m_e=1.0\,{\rm MeV}$ is the
last to depart the equilibrium track as the higher plasma temperature enhances
the neutron-to-proton rates.  Note that $(n/p)^{\rm (eq)}$ depends on the
evolution of $T$ which is different for each mass case.  However, the
differences are small at the level of precision of Fig.\ \ref{fig:np_vs_tcm}.
The changes in the out-of-equilibrium evolution of $n/p$ are much starker.

\begin{figure}
  \includegraphics[width=\columnwidth]{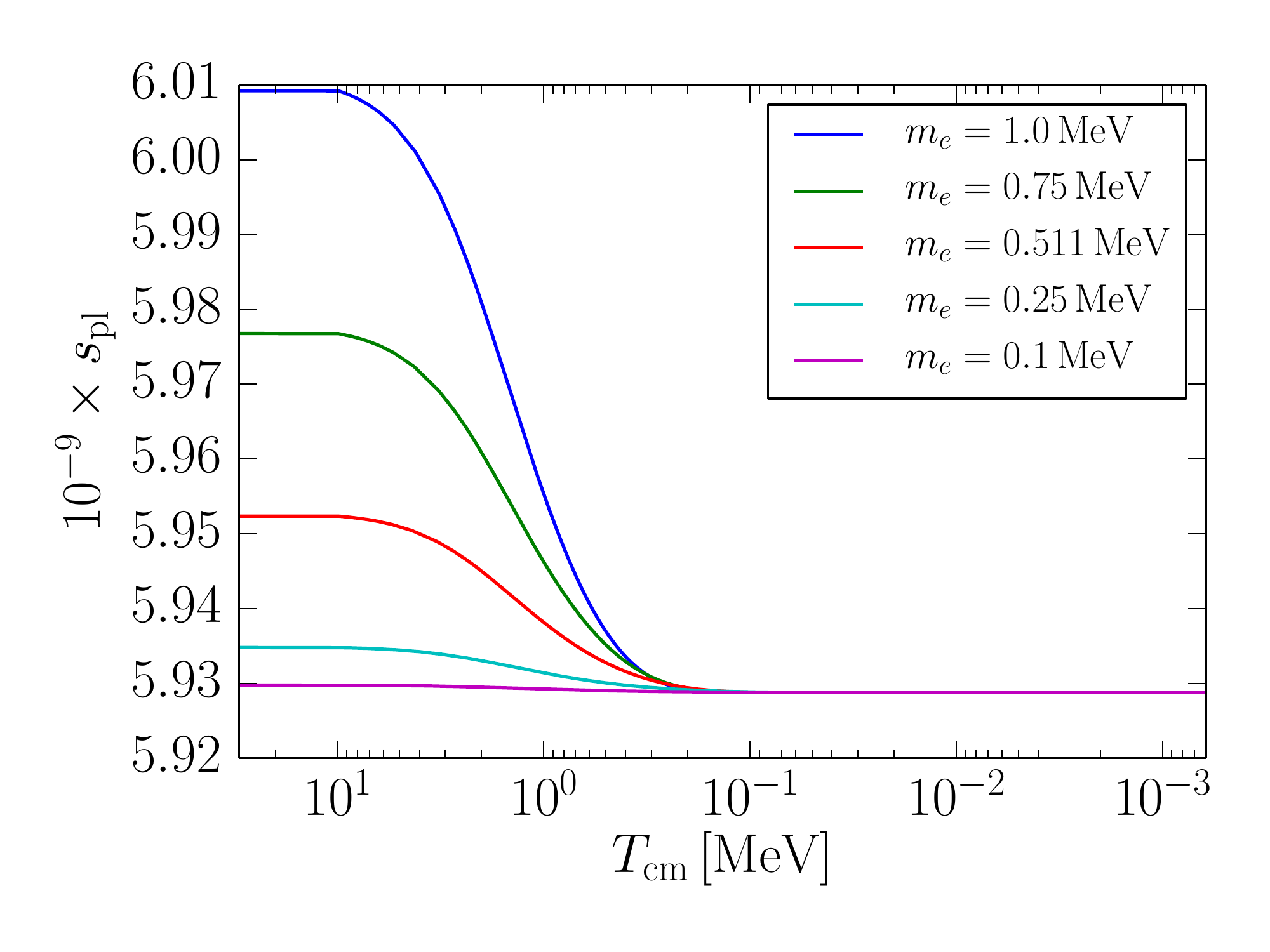}
  \caption{\label{fig:s_vs_tcm} Entropy per baryon in the plasma
  as a function of \tcm.}
\end{figure}

Figure \ref{fig:s_vs_tcm} shows the evolution of the entropy per baryon in the
plasma versus the comoving temperature parameter.  Electrons can annihilate with
positrons to produce neutrino-antineutrino pairs.  If the charged leptons have
larger masses, then each such annihilation event will produce more energetic neutrinos,
enabling a larger entropy transfer from the plasma into the neutrino seas.
Conversely, there are fewer $e^\pm$ pairs in equilibrium for
larger-mass charged leptons, implying fewer total annihilation events and a
smaller flow of entropy from the plasma into the neutrino seas.  Figure
\ref{fig:s_vs_tcm} clearly shows that the first effect dominates over the
second.

Two pieces of evidence support this result.  First, weak decoupling involves
the competition between the weak interaction rates and the Hubble expansion
rate.  We neither changed $G_F$ (the weak coupling constant) nor \mpl (the
Planck mass) implying that weak decoupling will occur roughly at the same
time/\tcm for different masses, as verified by Fig.\ \ref{fig:s_vs_tcm}.  It is
true that the dynamics of weak decoupling depend on the electron rest mass
through the pair density.  This leads to the second piece of evidence: weak
decoupling occurs during pair domination.  This is supported by Fig.\
\ref{fig:nep_vs_tcm}, which shows that even for $m_e=1.0\,{\rm MeV}$ there are
orders of magnitude more pairs than ionization electrons during the range of
\tcm particular to weak decoupling.  Therefore, the location of the entropy
flow is independent of the rest mass.  The magnitude of the flow increases with
increasing rest mass.  To precipitate an earlier epoch for the entropy flow,
the rest mass would need to be larger than $10\,{\rm MeV}$ so that neutrinos
would fall out of equilibrium because of a lack of scattering targets and not
from the low strength of the weak interaction.  Electron masses that large
would dictate a nonperturbative treatment incongruent with sections
\ref{sec:notrans} and \ref{sec:qed}.

\section{Conclusion}\label{sec:concl}

In this paper we have examined the role of charged lepton mass in the epoch of
weak decoupling and nucleosynthesis in the early universe.  We unphysically
vary the input vacuum electron rest mass as a means to dissect the complicated
and coupled nonlinear physics in this epoch.  Our goal was to gain deeper and
finer-scale insights into the physics of this epoch, with a subsidiary goal to
identify potential problems in high-precision calculations.  Clearly, the
subject of neutrino decoupling and BBN is quite correctly regarded as well
understood, with the basic ideas and calculations in place since the time of
Ref.\ \cite{WFH:1967} (see Ref.\ \cite{Cyburt:2016RMP} for a detailed review).

However, there is a renewed push for higher precision in calculations, driven
by the prospect of higher precision in future CMB experiments, i.e., CMB
Stage-IV \cite{cmbs4_science_book}, and the advent of 30-m class telescopes
with the possibility that we can obtain higher-precision primordial deuterium
measurements \cite{2016ApJ...830..148C}.  If the uncertainty in the
observational value of \neff can be reduced to less than $1\%$, then the
predicted value of $\neff=3.046$ (in line with our value $3.044$) would be
statistically different from $3.0$ at the $2\sigma$ level. Moreover, ideas for
beyond standard model (BSM) physics that might subtly alter the physics of this
epoch are ubiquitous
\cite{2006PhRvD..74j3509J,2014PhRvD..90c5022F,2014arXiv1411.1071C,
2016JCAP...01..007B}.  As an example, BSM physics which contains nonstandard
neutrino-electron interactions are sensitive to the dynamics of electrons and
positrons \cite{2016JCAP...07..051D,2006NuPhB.756..100M}. For these reasons, it
is important to understand the outstanding issues in the calculations that
impede better computational precision. Artificially changing the electron mass
and examining the consequences allowed us to leverage the overriding physical
facts of the early universe, large entropy and slow expansion rate, into a
deeper understanding of the interplay of the weak interaction and
thermodynamics at this epoch. This study has also helped to underline the
importance of an improved treatment of in-medium corrections to electron and
positron masses.

The electron rest mass directly plays a role in the thermodynamics, neutrino
energy transport, and neutron-to-proton weak-interaction rates.  Our
calculations show how $e^\pm$ pairs dominate over ionization electrons even at the low
energy scales expected at the end of the decoupling and BBN epoch.  There are two key
implications of the prolonged epoch of pair existence. First, there are more
charged-lepton targets for neutrino scattering at lower temperatures, implying
more transfer of entropy from the electromagnetic plasma to the decoupling
neutrinos.  The result of the entropy flow is an altered phasing of plasma
temperature with scale factor, as well as nonthermal distortions in the
decoupling neutrino energy spectra and a concomitant alteration to the ratio
\tcmplfo.  Nonzero electron rest mass, finite-temperature QED effects, and
out-of-equilibrium neutrino transport all increase the temperature ratio from
the standard equilibrium value of \felev.   Second, at lower temperature scales
more positrons induce an enhanced destruction of neutrons through the
no-threshold, lepton-capture process $e^++n\rightarrow p+\bnue$. The enhanced
destruction of neutrons alters the primordial helium yield, \yp, and the
primordial deuterium yield.

The third, fourth, and fifth columns of Table \ref{tab:trans_runs} address the
way in which three unique cosmological observables change with varying electron
rest mass.  These are the baryon density, \bardens, a measure of relativistic
energy density, \neff, and the primordial mass fraction of helium, \yp.  For
\bardens and \neff, the changes range from a few tenths of a percent to ten
percent.  The changes in the primordial helium abundance are much more drastic;
\yp changes at the $\sim\pm50\%$ level over the range of $m_e$ considered.
Most intriguing, even small corrections to in-medium electron mass produce
potentially observable nucleosynthesis effects.  For example, if we perturb
$m_e$ by $1\%$ from the true vacuum value (in the range of finite-temperature
QED corrections), we find changes in \yp at the $0.7\%$ level.  This is indeed
a small change, but Fig.\ \ref{fig:t_vs_xm_dme2me} showed that $\delta m_e^2$
can be as large as 10\% of $m_e^2$ at $T\sim1\,{\rm MeV}$.  Future CMB
experiments \cite{cmbs4_science_book} will achieve 1\% precision in
cosmological observables such as \neff and \yp.  Our work shows that
nucleosynthesis considerations (\yp and possibly deuterium) has the potential
to break degeneracies in beyond-standard-model physics.  Insights into neutrino
physics gleaned from nucleosynthesis would be complementary to those of \neff.
The frontier of precision in weak-decoupling-nucleosynthesis calculations lies
in an accurate treatment of neutrino physics, including neutrino-electron
scattering and ultimately neutrino flavor quantum kinetics.  Our work shows how
neutrino physics is tightly coupled within the physics of the electromagnetic
plasma, and thereby underscores the looming importance of improved plasma and
QED corrections to charge-lepton properties.

\section*{Acknowledgments}

We thank Fred Adams, Vincenzo Cirigliano, Chad Kishimoto, and Mark Paris for
useful conversations.  This work was supported in part by the Los Alamos
National Laboratory Institutional Computing Program, under U.S.\ Department of
Energy National Nuclear Security Administration Award No.\ DE-AC52-06NA25396;
and NSF grants PHY-1307372 and PHY-1614864 at UC San Diego.  We thank the
referee for their comments.


\bibliographystyle{elsarticle-num}
\bibliography{master}

\begin{thebibliography}{10}
\expandafter\ifx\csname url\endcsname\relax
  \def\url#1{\texttt{#1}}\fi
\expandafter\ifx\csname urlprefix\endcsname\relax\def\urlprefix{URL }\fi
\expandafter\ifx\csname href\endcsname\relax
  \def\href#1#2{#2} \def\path#1{#1}\fi

\bibitem{2014arXiv1412.1078S}
M.~{Shimon}, {A Globally Unevolving Universe}, ArXiv e-prints\href
  {http://arxiv.org/abs/1412.1078} {\path{arXiv:1412.1078}}.

\bibitem{PlanckXIII:2015}
{Planck Collaboration}, P.~A.~R. {Ade}, N.~{Aghanim}, M.~{Arnaud},
  M.~{Ashdown}, J.~{Aumont}, C.~{Baccigalupi}, A.~J. {Banday}, R.~B.
  {Barreiro}, J.~G. {Bartlett}, et~al., {Planck 2015 results. XIII.
  Cosmological parameters}, \aap 594 (2016) A13.
\newblock \href {http://arxiv.org/abs/1502.01589} {\path{arXiv:1502.01589}},
  \href {http://dx.doi.org/10.1051/0004-6361/201525830}
  {\path{doi:10.1051/0004-6361/201525830}}.

\bibitem{2003ApJS..149....1K}
D.~{Kirkman}, D.~{Tytler}, N.~{Suzuki}, J.~M. {O'Meara}, D.~{Lubin}, {The
  Cosmological Baryon Density from the Deuterium-to-Hydrogen Ratio in QSO
  Absorption Systems: D/H toward Q1243+3047}, \apjs 149 (2003) 1--28.
\newblock \href {http://arxiv.org/abs/astro-ph/0302006}
  {\path{arXiv:astro-ph/0302006}}, \href {http://dx.doi.org/10.1086/378152}
  {\path{doi:10.1086/378152}}.

\bibitem{2016ApJ...830..148C}
R.~J. {Cooke}, M.~{Pettini}, K.~M. {Nollett}, R.~{Jorgenson}, {The Primordial
  Deuterium Abundance of the Most Metal-poor Damped Lyman-{$\alpha$} System},
  \apj 830 (2016) 148.
\newblock \href {http://arxiv.org/abs/1607.03900} {\path{arXiv:1607.03900}},
  \href {http://dx.doi.org/10.3847/0004-637X/830/2/148}
  {\path{doi:10.3847/0004-637X/830/2/148}}.

\bibitem{1982NuPhB.209..372C}
J.-L. {Cambier}, J.~R. {Primack}, M.~{Sher}, {Finite temperature radiative
  corrections to neutron decay and related processes}, Nucl. Phys. B 209 (1982)
  372--388.
\newblock \href {http://dx.doi.org/10.1016/0550-3213(82)90262-0}
  {\path{doi:10.1016/0550-3213(82)90262-0}}.

\bibitem{1994PhRvD..49..611H}
A.~F. {Heckler}, {Astrophysical applications of quantum corrections to the
  equation of state of a plasma}, \prd 49 (1994) 611--617.
\newblock \href {http://dx.doi.org/10.1103/PhysRevD.49.611}
  {\path{doi:10.1103/PhysRevD.49.611}}.

\bibitem{1997PhRvD..56.5123F}
N.~{Fornengo}, C.~W. {Kim}, J.~{Song}, {Finite temperature effects on the
  neutrino decoupling in the early Universe}, \prd 56 (1997) 5123--5134.
\newblock \href {http://arxiv.org/abs/hep-ph/9702324}
  {\path{arXiv:hep-ph/9702324}}, \href
  {http://dx.doi.org/10.1103/PhysRevD.56.5123}
  {\path{doi:10.1103/PhysRevD.56.5123}}.

\bibitem{1999PhRvD..59j3502L}
R.~E. {Lopez}, M.~S. {Turner}, {Precision prediction for the big-bang abundance
  of primordial $^{4}$He}, \prd 59~(10) (1999) 103502.
\newblock \href {http://arxiv.org/abs/astro-ph/9807279}
  {\path{arXiv:astro-ph/9807279}}, \href
  {http://dx.doi.org/10.1103/PhysRevD.59.103502}
  {\path{doi:10.1103/PhysRevD.59.103502}}.

\bibitem{neff:3.046}
G.~{Mangano}, G.~{Miele}, S.~{Pastor}, T.~{Pinto}, O.~{Pisanti}, P.~D.
  {Serpico}, {Relic neutrino decoupling including flavour oscillations},
  Nuclear Physics B 729 (2005) 221--234.
\newblock \href {http://arxiv.org/abs/hep-ph/0506164}
  {\path{arXiv:hep-ph/0506164}}, \href
  {http://dx.doi.org/10.1016/j.nuclphysb.2005.09.041}
  {\path{doi:10.1016/j.nuclphysb.2005.09.041}}.

\bibitem{2008CoPhC.178..956P}
O.~{Pisanti}, A.~{Cirillo}, S.~{Esposito}, F.~{Iocco}, G.~{Mangano},
  G.~{Miele}, P.~D. {Serpico}, {PArthENoPE: Public algorithm evaluating the
  nucleosynthesis of primordial elements}, Computer Physics Communications 178
  (2008) 956--971.
\newblock \href {http://arxiv.org/abs/0705.0290} {\path{arXiv:0705.0290}},
  \href {http://dx.doi.org/10.1016/j.cpc.2008.02.015}
  {\path{doi:10.1016/j.cpc.2008.02.015}}.

\bibitem{transport_paper}
E.~{Grohs}, G.~M. {Fuller}, C.~T. {Kishimoto}, M.~W. {Paris}, A.~{Vlasenko},
  {Neutrino energy transport in weak decoupling and big bang nucleosynthesis},
  \prd 93~(8) (2016) 083522.
\newblock \href {http://arxiv.org/abs/1512.02205} {\path{arXiv:1512.02205}},
  \href {http://dx.doi.org/10.1103/PhysRevD.93.083522}
  {\path{doi:10.1103/PhysRevD.93.083522}}.

\bibitem{1990eaun.book.....K}
E.~W. {Kolb}, M.~S. {Turner}, {The Early Universe.}, Addison-Wesley Publishing
  Co., 1990.

\bibitem{Dodelson:2003mc}
S.~Dodelson, \href{http://books.google.com/books?id=3oPRxdXJexcC}{Modern
  Cosmology}, Academic Press, Academic Press, 2003.
\newline\urlprefix\url{http://books.google.com/books?id=3oPRxdXJexcC}

\bibitem{2008cosm.book.....W}
S.~{Weinberg}, {Cosmology}, Oxford University Press, 2008.

\bibitem{Pathria_stat_mech}
R.~K. Press, Statistical Mechanics, 2nd Edition, Butterworth-Heinemann, Woburn,
  MA, USA, 1996.

\bibitem{PlanckXVI:2014}
{Planck Collaboration}, P.~A.~R. {Ade}, et~al., {Planck 2013 results. XVI.
  Cosmological parameters}, \aap 571 (2014) A16.
\newblock \href {http://arxiv.org/abs/1303.5076} {\path{arXiv:1303.5076}},
  \href {http://dx.doi.org/10.1051/0004-6361/201321591}
  {\path{doi:10.1051/0004-6361/201321591}}.

\bibitem{Mangano:3.040}
G.~{Mangano}, G.~{Miele}, S.~{Pastor}, M.~{Peloso}, {A precision calculation of
  the effective number of cosmological neutrinos}, Physics Letters B 534 (2002)
  8--16.
\newblock \href {http://arxiv.org/abs/astro-ph/0111408}
  {\path{arXiv:astro-ph/0111408}}, \href
  {http://dx.doi.org/10.1016/S0370-2693(02)01622-2}
  {\path{doi:10.1016/S0370-2693(02)01622-2}}.

\bibitem{Blaschke_Cirigliano_2016}
D.~N. {Blaschke}, V.~{Cirigliano}, {Neutrino quantum kinetic equations: The
  collision term}, \prd 94~(3) (2016) 033009.
\newblock \href {http://arxiv.org/abs/1605.09383} {\path{arXiv:1605.09383}},
  \href {http://dx.doi.org/10.1103/PhysRevD.94.033009}
  {\path{doi:10.1103/PhysRevD.94.033009}}.

\bibitem{1991NuPhB.349..743B}
R.~{Barbieri}, A.~{Dolgov}, {Neutrino oscillations in the early universe},
  Nuclear Physics B 349 (1991) 743--753.
\newblock \href {http://dx.doi.org/10.1016/0550-3213(91)90396-F}
  {\path{doi:10.1016/0550-3213(91)90396-F}}.

\bibitem{AkhmedovBerezhiani}
E.~K. {Akhmedov}, Z.~G. {Berezhiani}, {Implications of Majorana neutrino
  transition magnetic moments for neutrino signals from supernovae}, Nuclear
  Physics B 373 (1992) 479--497.
\newblock \href {http://dx.doi.org/10.1016/0550-3213(92)90441-D}
  {\path{doi:10.1016/0550-3213(92)90441-D}}.

\bibitem{1993APh.....1..165R}
G.~{Raffelt}, G.~{Sigl}, {Neutrino flavor conversion in a supernova core},
  Astroparticle Physics 1 (1993) 165--183.
\newblock \href {http://arxiv.org/abs/astro-ph/9209005}
  {\path{arXiv:astro-ph/9209005}}, \href
  {http://dx.doi.org/10.1016/0927-6505(93)90020-E}
  {\path{doi:10.1016/0927-6505(93)90020-E}}.

\bibitem{2005PhRvD..71i3004S}
P.~{Strack}, A.~{Burrows}, {Generalized Boltzmann formalism for oscillating
  neutrinos}, \prd 71~(9) (2005) 093004.
\newblock \href {http://arxiv.org/abs/hep-ph/0504035}
  {\path{arXiv:hep-ph/0504035}}, \href
  {http://dx.doi.org/10.1103/PhysRevD.71.093004}
  {\path{doi:10.1103/PhysRevD.71.093004}}.

\bibitem{2007JPhG...34...47B}
A.~B. {Balantekin}, Y.~{Pehlivan}, {Neutrino neutrino interactions and flavour
  mixing in dense matter}, Journal of Physics G Nuclear Physics 34 (2007)
  47--65.
\newblock \href {http://arxiv.org/abs/astro-ph/0607527}
  {\path{arXiv:astro-ph/0607527}}, \href
  {http://dx.doi.org/10.1088/0954-3899/34/1/004}
  {\path{doi:10.1088/0954-3899/34/1/004}}.

\bibitem{2013PhRvD..87k3010V}
C.~{Volpe}, D.~{V{\"a}{\"a}n{\"a}nen}, C.~{Espinoza}, {Extended evolution
  equations for neutrino propagation in astrophysical and cosmological
  environments}, \prd 87~(11) (2013) 113010.
\newblock \href {http://arxiv.org/abs/1302.2374} {\path{arXiv:1302.2374}},
  \href {http://dx.doi.org/10.1103/PhysRevD.87.113010}
  {\path{doi:10.1103/PhysRevD.87.113010}}.

\bibitem{2013PrPNP..71..162B}
A.~B. {Balantekin}, G.~M. {Fuller}, {Neutrinos in cosmology and astrophysics},
  Progress in Particle and Nuclear Physics 71 (2013) 162--166.
\newblock \href {http://arxiv.org/abs/1303.3874} {\path{arXiv:1303.3874}},
  \href {http://dx.doi.org/10.1016/j.ppnp.2013.03.008}
  {\path{doi:10.1016/j.ppnp.2013.03.008}}.

\bibitem{Gouvea}
A.~{de Gouv{\^e}a}, S.~{Shalgar}, {Transition magnetic moments and collective
  neutrino oscillations: three-flavor effects and detectability}, \jcap 4
  (2013) 18.
\newblock \href {http://arxiv.org/abs/1301.5637} {\path{arXiv:1301.5637}},
  \href {http://dx.doi.org/10.1088/1475-7516/2013/04/018}
  {\path{doi:10.1088/1475-7516/2013/04/018}}.

\bibitem{VFC:QKE}
A.~{Vlasenko}, G.~M. {Fuller}, V.~{Cirigliano}, {Neutrino quantum kinetics},
  \prd 89~(10) (2014) 105004.
\newblock \href {http://arxiv.org/abs/1309.2628} {\path{arXiv:1309.2628}},
  \href {http://dx.doi.org/10.1103/PhysRevD.89.105004}
  {\path{doi:10.1103/PhysRevD.89.105004}}.

\bibitem{2014PhRvD..90l5040S}
J.~{Serreau}, C.~{Volpe}, {Neutrino-antineutrino correlations in dense
  anisotropic media}, \prd 90~(12) (2014) 125040.
\newblock \href {http://arxiv.org/abs/1409.3591} {\path{arXiv:1409.3591}},
  \href {http://dx.doi.org/10.1103/PhysRevD.90.125040}
  {\path{doi:10.1103/PhysRevD.90.125040}}.

\bibitem{2015PhRvD..91l5020K}
A.~{Kartavtsev}, G.~{Raffelt}, H.~{Vogel}, {Neutrino propagation in media:
  Flavor, helicity, and pair correlations}, \prd 91~(12) (2015) 125020.
\newblock \href {http://arxiv.org/abs/1504.03230} {\path{arXiv:1504.03230}},
  \href {http://dx.doi.org/10.1103/PhysRevD.91.125020}
  {\path{doi:10.1103/PhysRevD.91.125020}}.

\bibitem{2015PhLB..747...27C}
V.~{Cirigliano}, G.~M. {Fuller}, A.~{Vlasenko}, {A new spin on neutrino quantum
  kinetics}, Physics Letters B 747 (2015) 27--35.
\newblock \href {http://arxiv.org/abs/1406.5558} {\path{arXiv:1406.5558}},
  \href {http://dx.doi.org/10.1016/j.physletb.2015.04.066}
  {\path{doi:10.1016/j.physletb.2015.04.066}}.

\bibitem{2015PhyS...90h8008K}
B.~D. {Keister}, {Numerical and physical stability of supernova neutrino flavor
  evolution}, \physscr 90~(8) (2015) 088008.
\newblock \href {http://arxiv.org/abs/1408.4729} {\path{arXiv:1408.4729}},
  \href {http://dx.doi.org/10.1088/0031-8949/90/8/088008}
  {\path{doi:10.1088/0031-8949/90/8/088008}}.

\bibitem{2015IJMPE..2441009V}
C.~{Volpe}, {Neutrino quantum kinetic equations}, International Journal of
  Modern Physics E 24 (2015) 1541009.
\newblock \href {http://arxiv.org/abs/1506.06222} {\path{arXiv:1506.06222}},
  \href {http://dx.doi.org/10.1142/S0218301315410098}
  {\path{doi:10.1142/S0218301315410098}}.

\bibitem{2016NuPhB.908..366C}
S.~{Chakraborty}, R.~{Hansen}, I.~{Izaguirre}, G.~{Raffelt}, {Collective
  neutrino flavor conversion: Recent developments}, Nuclear Physics B 908
  (2016) 366--381.
\newblock \href {http://arxiv.org/abs/1602.02766} {\path{arXiv:1602.02766}},
  \href {http://dx.doi.org/10.1016/j.nuclphysb.2016.02.012}
  {\path{doi:10.1016/j.nuclphysb.2016.02.012}}.

\bibitem{2016JCAP...12..019S}
R.~{Sloth Lundkvist Hansen}, A.~Y. {Smirnov}, {The Liouville equation for
  flavour evolution of neutrinos and neutrino wave packets}, \jcap 12 (2016)
  019.
\newblock \href {http://arxiv.org/abs/1610.00910} {\path{arXiv:1610.00910}},
  \href {http://dx.doi.org/10.1088/1475-7516/2016/12/019}
  {\path{doi:10.1088/1475-7516/2016/12/019}}.

\bibitem{1993NuPhB.406..423S}
G.~{Sigl}, G.~{Raffelt}, {General kinetic description of relativistic mixed
  neutrinos}, Nuclear Physics B 406 (1993) 423--451.
\newblock \href {http://dx.doi.org/10.1016/0550-3213(93)90175-O}
  {\path{doi:10.1016/0550-3213(93)90175-O}}.

\bibitem{1995PhRvD..52.1764H}
S.~{Hannestad}, J.~{Madsen}, {Neutrino decoupling in the early Universe}, \prd
  52 (1995) 1764--1769.
\newblock \href {http://arxiv.org/abs/astro-ph/9506015}
  {\path{arXiv:astro-ph/9506015}}, \href
  {http://dx.doi.org/10.1103/PhysRevD.52.1764}
  {\path{doi:10.1103/PhysRevD.52.1764}}.

\bibitem{Dolgov:1997ne}
A.~D. {Dolgov}, S.~H. {Hansen}, D.~V. {Semikoz}, {Non-equilibrium corrections
  to the spectra of massless neutrinos in the early universe}, Nuclear Physics
  B 503 (1997) 426--444.
\newblock \href {http://arxiv.org/abs/hep-ph/9703315}
  {\path{arXiv:hep-ph/9703315}}, \href
  {http://dx.doi.org/10.1016/S0550-3213(97)00479-3}
  {\path{doi:10.1016/S0550-3213(97)00479-3}}.

\bibitem{letsgoeu2}
L.~{Kawano}, {Let's go: Early universe 2. Primordial nucleosynthesis the
  computer way}, NASA STI/Recon Technical Report 92 (1992) 25163.

\bibitem{lep_trans}
E.~{Grohs}, G.~M. {Fuller}, C.~T. {Kishimoto}, M.~W. {Paris}, {Lepton
  asymmetry, neutrino spectral distortions, and big bang nucleosynthesis}, \prd
  95~(6) (2017) 063503.
\newblock \href {http://arxiv.org/abs/1612.01986} {\path{arXiv:1612.01986}},
  \href {http://dx.doi.org/10.1103/PhysRevD.95.063503}
  {\path{doi:10.1103/PhysRevD.95.063503}}.

\bibitem{2016JCAP...07..051D}
P.~F. {de Salas}, S.~{Pastor}, {Relic neutrino decoupling with flavour
  oscillations revisited}, \jcap 7 (2016) 051.
\newblock \href {http://arxiv.org/abs/1606.06986} {\path{arXiv:1606.06986}},
  \href {http://dx.doi.org/10.1088/1475-7516/2016/07/051}
  {\path{doi:10.1088/1475-7516/2016/07/051}}.

\bibitem{2015NuPhB.890..481B}
J.~{Birrell}, C.~T. {Yang}, J.~{Rafelski}, {Relic neutrino freeze-out:
  Dependence on natural constants}, Nuclear Physics B 890 (2015) 481--517.
\newblock \href {http://arxiv.org/abs/1406.1759} {\path{arXiv:1406.1759}},
  \href {http://dx.doi.org/10.1016/j.nuclphysb.2014.11.020}
  {\path{doi:10.1016/j.nuclphysb.2014.11.020}}.

\bibitem{1982PhRvD..26.2694D}
D.~A. {Dicus}, E.~W. {Kolb}, A.~M. {Gleeson}, E.~C.~G. {Sudarshan}, V.~L.
  {Teplitz}, M.~S. {Turner}, {Primordial nucleosynthesis including radiative,
  Coulomb, and finite-temperature corrections to weak rates}, \prd 26 (1982)
  2694--2706.
\newblock \href {http://dx.doi.org/10.1103/PhysRevD.26.2694}
  {\path{doi:10.1103/PhysRevD.26.2694}}.

\bibitem{FFN:I}
G.~M. {Fuller}, W.~A. {Fowler}, M.~J. {Newman}, {Stellar weak-interaction rates
  for sd-shell nuclei. I - Nuclear matrix element systematics with application
  to Al-26 and selected nuclei of importance to the supernova problem}, \apjs
  42 (1980) 447--473.
\newblock \href {http://dx.doi.org/10.1086/190657} {\path{doi:10.1086/190657}}.

\bibitem{FFN:III}
G.~M. {Fuller}, W.~A. {Fowler}, M.~J. {Newman}, {Stellar weak interaction rates
  for intermediate mass nuclei. III - Rate tables for the free nucleons and
  nuclei with A = 21 to A = 60}, \apjs 48 (1982) 279--319.
\newblock \href {http://dx.doi.org/10.1086/190779} {\path{doi:10.1086/190779}}.

\bibitem{FFN:IV}
G.~M. {Fuller}, W.~A. {Fowler}, M.~J. {Newman}, {Stellar weak interaction rates
  for intermediate-mass nuclei. IV - Interpolation procedures for rapidly
  varying lepton capture rates using effective log (ft)-values}, \apj 293
  (1985) 1--16.
\newblock \href {http://dx.doi.org/10.1086/163208} {\path{doi:10.1086/163208}}.

\bibitem{WFO_approx}
E.~{Grohs}, G.~M. {Fuller}, {The surprising influence of late charged current
  weak interactions on Big Bang Nucleosynthesis}, Nuclear Physics B 911 (2016)
  955--973.
\newblock \href {http://arxiv.org/abs/1607.02797} {\path{arXiv:1607.02797}},
  \href {http://dx.doi.org/10.1016/j.nuclphysb.2016.08.034}
  {\path{doi:10.1016/j.nuclphysb.2016.08.034}}.

\bibitem{WFH:1967}
R.~V. Wagoner, W.~A. Fowler, F.~Hoyle, {On the Synthesis of elements at very
  high temperatures}, Astrophys.J. 148 (1967) 3--49.
\newblock \href {http://dx.doi.org/10.1086/149126} {\path{doi:10.1086/149126}}.

\bibitem{Cyburt:2016RMP}
R.~H. {Cyburt}, B.~D. {Fields}, K.~A. {Olive}, T.-H. {Yeh}, {Big bang
  nucleosynthesis: Present status}, Reviews of Modern Physics 88~(1) (2016)
  015004.
\newblock \href {http://arxiv.org/abs/1505.01076} {\path{arXiv:1505.01076}},
  \href {http://dx.doi.org/10.1103/RevModPhys.88.015004}
  {\path{doi:10.1103/RevModPhys.88.015004}}.

\bibitem{cmbs4_science_book}
J.~E. {Carlstrom}, et~al., {CMB-S4 Science Book, First Edition}, ArXiv
  e-prints\href {http://arxiv.org/abs/1610.02743} {\path{arXiv:1610.02743}}.

\bibitem{2006PhRvD..74j3509J}
K.~{Jedamzik}, {Big bang nucleosynthesis constraints on hadronically and
  electromagnetically decaying relic neutral particles}, \prd 74~(10) (2006)
  103509.
\newblock \href {http://arxiv.org/abs/hep-ph/0604251}
  {\path{arXiv:hep-ph/0604251}}, \href
  {http://dx.doi.org/10.1103/PhysRevD.74.103509}
  {\path{doi:10.1103/PhysRevD.74.103509}}.

\bibitem{2014PhRvD..90c5022F}
A.~{Fradette}, M.~{Pospelov}, J.~{Pradler}, A.~{Ritz}, {Cosmological
  constraints on very dark photons}, \prd 90~(3) (2014) 035022.
\newblock \href {http://arxiv.org/abs/1407.0993} {\path{arXiv:1407.0993}},
  \href {http://dx.doi.org/10.1103/PhysRevD.90.035022}
  {\path{doi:10.1103/PhysRevD.90.035022}}.

\bibitem{2014arXiv1411.1071C}
J.~F. {Cherry}, A.~{Friedland}, I.~M. {Shoemaker}, {Neutrino Portal Dark
  Matter: From Dwarf Galaxies to IceCube}, ArXiv e-prints\href
  {http://arxiv.org/abs/1411.1071} {\path{arXiv:1411.1071}}.

\bibitem{2016JCAP...01..007B}
D.~{Baumann}, D.~{Green}, J.~{Meyers}, B.~{Wallisch}, {Phases of new physics in
  the CMB}, \jcap 1 (2016) 007.
\newblock \href {http://arxiv.org/abs/1508.06342} {\path{arXiv:1508.06342}},
  \href {http://dx.doi.org/10.1088/1475-7516/2016/01/007}
  {\path{doi:10.1088/1475-7516/2016/01/007}}.

\bibitem{2006NuPhB.756..100M}
G.~{Mangano}, G.~{Miele}, S.~{Pastor}, T.~{Pinto}, O.~{Pisanti}, P.~D.
  {Serpico}, {Effects of non-standard neutrino electron interactions on relic
  neutrino decoupling}, Nuclear Physics B 756 (2006) 100--116.
\newblock \href {http://arxiv.org/abs/hep-ph/0607267}
  {\path{arXiv:hep-ph/0607267}}, \href
  {http://dx.doi.org/10.1016/j.nuclphysb.2006.09.002}
  {\path{doi:10.1016/j.nuclphysb.2006.09.002}}.

\end{thebibliography}

\end{document}